\documentclass[final]{elsarticle}
\pdfoutput=1

\usepackage{lineno,hyperref}
\usepackage[numbers]{natbib}
\usepackage{amsmath} 
\usepackage{amssymb}  
\usepackage{bm}
\usepackage{graphicx}
\usepackage{subcaption}
\usepackage{float}
\usepackage{enumitem}
\usepackage{color,soul}
\usepackage{array, makecell}
\usepackage{xcolor}
\usepackage{algorithm}
\usepackage{hyperref}
\modulolinenumbers[5]

\journal{Journal of Mechanical Systems and Signal Processing}

\begin{document}

\begin{frontmatter}

\title{Applying Polynomial Decoupling Methods to the Polynomial NARX Model  \tnoteref{mytitlenote}}
\tnotetext[mytitlenote]{This work is supported by the Natural Sciences and Engineering Research Council of Canada (NSERC) through grant RGPIN/06464-2015 and also the Fund for Scientific Research FWO project G0068.18N.}

\author[1]{Kiana Karami\corref{mycorrespondingauthor}}
\cortext[mycorrespondingauthor]{Corresponding author, Tel. +1-587-969-4940, Fax +1-403-282-6855}
\ead{kiana.karami@ucalgary.ca}

\author[1]{David Westwick}
\ead{dwestwic@ucalgary.ca}

\author[2,3]{Johan Schoukens}
\ead{johan.Schoukens@vub.be}

\address[1]{Department of Electrical and Computer Engineering, Schulich School of Engineering, University of Calgary, 2500 University Drive NW, Calgary, Alberta, T2N 1N4, Canada. }
\address[2]{Department INDI, Vrije Universiteit Brussel, Boulevard de la Plaine 2, 1050 Ixelles, Belgium.}            
\address[3]{Department of Electrical Engineering, Eindhoven University of Technology, Groene Loper 19, 5612 AP Eindhoven, Netherlands.}

\begin{abstract}
System identification uses measurements of a dynamic system's input and output to reconstruct a mathematical model for that system. These can be mechanical, electrical, physiological, among others.
Since most of the systems around us exhibit some form of nonlinear behavior, nonlinear system identification techniques are the tools that will help us gain a better understanding of our surroundings and potentially let us improve their performance. One model that is often used to represent nonlinear systems is the polynomial NARX model, an equation error model where the output is a polynomial function of the past inputs and outputs. That said, a major disadvantage with the polynomial NARX model is that the number of parameters increases rapidly with increasing polynomial order. Furthermore, the polynomial NARX model is a black-box model, and is therefore difficult to interpret.
This paper discusses a decoupling algorithm for the polynomial NARX model that substitutes the multivariate polynomial with a transformation matrix followed by a bank of univariate polynomials. This decreases the number of model parameters significantly and also imposes structure on the black-box NARX model. Since a non-convex optimization is required for this identification technique, initialization is an important factor to consider. In this paper the decoupling algorithm is developed in conjunction with several different initialization techniques. The resulting algorithms are applied to two nonlinear benchmark problems: measurement data from the Silver-Box and simulation data from the Bouc-Wen friction model, and the performance is evaluated for different validation signals in both simulation and prediction.
\end{abstract}

\begin{keyword}
system identification \sep nonlinear system \sep decoupled polynomial \sep NARX
\end{keyword}

\end{frontmatter}


\section{Introduction}
In the field of system identification, the goal is to build a mathematical model for a process or dynamic system  using measurements of its inputs and outputs that captures the relationship between these signals, within a reasonable degree of accuracy
\cite{ljung2001system,jategaonkar2004aerodynamic,soal2019system}.

Linear system identification methods have been studied extensively \cite{lennartsystem,pintelon2012system,schoukens2014system} and have been used in literature broadly \cite{Pintelon2010a,Pintelon2010b}. But most of the physical systems around us exhibit nonlinear behaviors. Although linear approximations of nonlinear systems have been used in many applications \cite{schoukens2016linear}, these models might not be accurate enough in some cases \cite{karami2015identification,ljung2010perspectives}, thus bringing up the need for nonlinear system identification methods \cite{schoukens2019nonlinear}.

The area of nonlinear system identification continues to be an active area of research \cite{kerschen2006past,noel2017nonlinear,lei2019novel,worden2018evolutionary}.
Indeed, this journal published a collection of papers  \cite{noel2019cross} based around a set of nonlinear system identification benchmark problems. The data from these benchmarks are freely available, and downloadable from  (\url{www.nonlinearbenchmark.org}), allowing researchers in this area to easily share their ideas and interact with each other. Data from two of these benchmark problems will be used to validate the algorithms developed in this paper.

A common nonlinear model that has been used extensively in system identification and modeling is the Nonlinear Auto-Regressive eXogenous Input (NARX) model. The NARX model is a simplification of the NARMAX model developed by Billings \cite{billings2013nonlinear} and many collaborators, and has been widely used in the literature \cite{caswell2014nonlinear,chan2015application,ruiz2016application,zhao2013new}.  This model relates the current value of the system's output to both past values of the output and current and past values of the input. In addition, the output equation contains an error term  which is assumed to be a sequence of independent identically distributed (IID) random variables. Such a model can be stated algebraically as:
\begin{align} \label{y-NARX}
 y(t) & = F(y(t-1), \cdots, y(t-n_y), \nonumber \\
 &\rule{0.9cm}{0cm} u(t),u(t-1),\cdots, u(t-n_u) )+e(t)
\end{align}
where $t$ is the discrete-time index, $u(t)$, $y(t)$ and $e(t)$ are the input, output and equation error, respectively, and $n_u$ and $n_y$ are the number of past input and output terms, respectively, included in the model. 

The function $F$ is a multivariate  nonlinear function that can be represented parametrically, such as with a spline \cite{shmilovici1999adaptive} or polynomial \cite{jones1989recursive,piroddi2003identification}, or  nonparametrically, for example using an artificial neural network \cite{alcan2019predicting} or a Gaussian Process model \cite{kocijan2005dynamic}.
When $F$ is a nonlinear polynomial function of its variables, this results in the Polynomial NARX (P-NARX) model.
In the P-NARX model, the output is a linear function of the parameters. Hence, the coefficients can be calculated using a linear regression. On the other hand, in general NARX models there is no guarantee that the output is linear in the variables, for example, in the case that $F$ is represented by a sigmoidal neural network, the output will depend nonlinearly on all of the weights and biases in the network.

Although the NARX model is widely used, it has major disadvantages such as: there are too many parameters to identify, and it is a black-box model so it is difficult to interpret, and not helpful if we are interested in system's structure or internal functions \cite{ljung2010perspectives}. 

One of the disadvantages of the P-NARX model is due to its structure, the number of parameters increases with the number of past inputs and outputs, and also increases combinatorally with the polynomial degree. Dreesen et al. \cite{dreesen2015decoupling} proposed a method for decoupling multiple-input multiple-output (MIMO) polynomial functions which has been used to simplify MIMO polynomials in various applications \cite{dreesen2015recovering,NoelSchoukens2017,esfahani2018parameter}. The MIMO polynomial is replaced by a bank of SISO polynomial functions, sandwiched between two linear transformation matrices. 
In \cite{dreesen2015decoupling}, a MIMO polynomial $ {\mathbf{y} } = f({\bf z})$, where ${\mathbf{y}} = [y_1, y_2, … y_p]^T$, was represented by the structure
$${\mathbf{y}} = {\mathbf{W}} g({\bm {V}}^T{\bf z})$$
\begin{figure}[h] 
\centering
\begin{subfigure}[b]{0.55\textwidth}
   \includegraphics[width = 80mm]{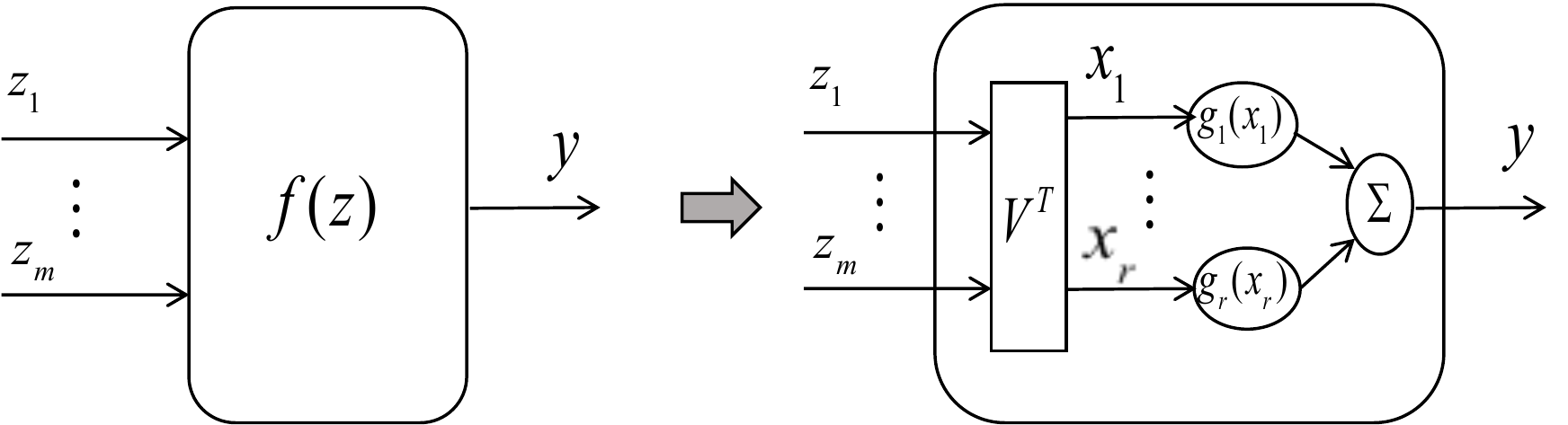}
   \caption{}
   \label{fig:decoupled MISO} 
\end{subfigure}

\begin{subfigure}[b]{0.55\textwidth}
   \includegraphics[width = 85mm]{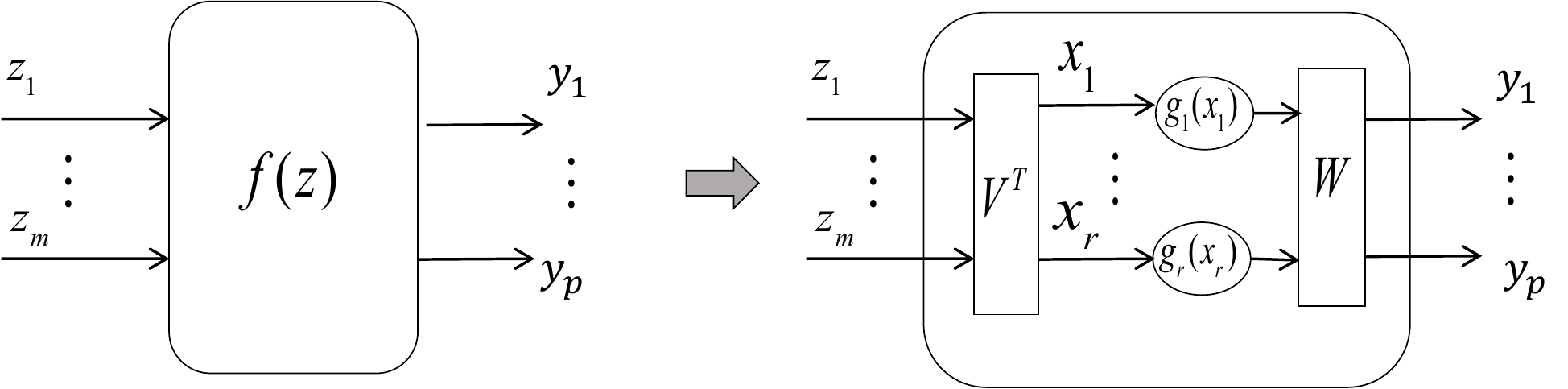}
   \caption{}
   \label{fig:decoupled MIMO}
\end{subfigure}

\caption[]{(a) Decoupling a MISO Nonlinearity Into SISO Polynomial Branches and a Transformation Matrix. (b)Decoupling a MIMO Polynomial as described in Dreesen et. al \cite{dreesen2015decoupling}
}\label{fig:decoupled} 
\end{figure}

Unlike the MIMO polynomials considered by Dreesen et al. \cite{dreesen2015decoupling}, the P-NARX model is a multiple-input, single-output (MISO) polynomial.  We introduced a decoupling algorithm \cite{westwick2018using} for MISO polynomials, and hence P-NARX models. Other techniques exist such as sparse regression, that create sparse models by forcing coefficients to be zero.
Decoupling is a 'milder' approach, where coefficients are not forced to zero. We are looking rather for a basis in which only a few coefficients are different from zero without forcing them to zero.

In the examples presented in \cite{westwick2018using}, decoupling the P-NARX model reduced the number of estimated parameters significantly. However, a major drawback to this method is that the model is nonlinear in some parameters. Hence a non-convex optimization is required for its identification. This paper further develops the method proposed in \cite{westwick2018using} and proposes some techniques that could be used to produce a good initial starting point for the optimization. In summary:
\begin{itemize}
    \item The final optimization used to fit the decoupled model is the same as proposed in \cite{westwick2018using}. This is a non-convex optimization, and
hence requires a good initialization to improve the chances of convergence to a good local minimum.
    \item In this paper, we introduce an initial solution to the decoupling problem, based on factoring the Hessian tensor of the initial, coupled
model, using Canonical Polyadic Decomposition with the addition of a polynomial constraint on one of the factors. Imposing a constraint on the CPD, ensures that the factor matrices can be generated by a decoupled polynomial structure.
    \item The possibility of using non-polynomial basis function has been discussed in this paper. The method provided in this paper can be extended to non-polynomial basis functions as long as they have a similar structure to the polynomials (multivariate basis elements are represented by products of single-variable terms).
    \item New results were generated on the benchmark data sets employing this initialization, and the results are compared to those
obtained using the approach proposed in \cite{westwick2018using}.
\end{itemize}

The balance of this paper is organized as follows, in Section \ref{DPNARX} the decoupled MISO polynomial NARX model will be introduced. Section \ref{opt-in} covers the optimization initialization process and Section \ref{alg} summarizes the proposed algorithm. Then, in Section \ref{benchmark} the proposed algorithm will be used to identify models of two benchmark problems: the Silver-Box and the Bouc-Wen. Finally, Section \ref{conc} will summarize the paper.

\subsection{Notation} 
Lower and upper case letters in a regular type-face, $a, A$, will refer to scalars, bold faced lower case letters, $\bm a$, refer to vectors, matrices are indicated by bold faced upper case letters, $\bm A$, and bold faced calligraphic script will be used for tensors, $\bm{\mathcal {A}}$. $\otimes$ will denote the Kronecker product.

\section{Decoupled Polynomial NARX Model} \label{DPNARX}
This section continues as follows; Section \ref{sec:DMulti} describes the general multivariate polynomial decoupling problem. Section \ref{Estimation of D model} describes an optimization solution to this problem. Finally, Section \ref{sec:extension} extends the polynomial decoupling method to NARX models.

Assume that $\mathbf{z} = [\begin{array}{cccc} z_1& z_2& \cdots& z_m \end{array}]^T$ contains the inputs to the polynomial as shown in Fig. \ref{fig:decoupled}.  
Then the polynomial can be written explicitly as:  
\begin{align*}
 F(z_1, \cdots, z_m) 
          = \sum_{\ell = 0}^{M} \rule{0.15cm}{0cm} \sum_{k_1 = 0}^{m} \rule{0.15cm}{0cm} \sum_{k_2 = k_1}^{m} \cdots \sum_{k_n = k_n-1}^{m} \rule{0.15cm}{0cm} 
              \rule{0.1cm}{0cm} \left[c_{\ell}(k_1,k_2, \cdots, k_{\ell})
\prod_{i=1}^{m} z_i  \right]
\end{align*}
where $c_{\ell}$ are the polynomial coefficients of degree ${\ell}$.
If the polynomial represents a P-NARX model, the $z_i$ will contain past inputs and outputs.

\subsection{Decoupled Multivariate Polynomial}\label{sec:DMulti}
The polynomial decoupling methods \cite{dreesen2015decoupling,westwick2018using} replace a multivariate polynomial that may include cross-terms involving several inputs with a bank of single-input single-output (SISO) polynomials.  In the MIMO case, considered in \cite{dreesen2015decoupling} and shown in Fig. \ref{fig:decoupled MIMO}, this bank of univariate polynomials was sandwiched between two linear transformation matrices.  In the MISO case \cite{westwick2018using}, the inputs to the univariate polynomials are linear combinations of the system's inputs, as in the MIMO case, but the outputs of the univariate polynomials are simply summed together (see Fig. \ref{fig:decoupled MISO}).

Thus, let $y({\mathbf{z}})$ 
be the output of a MISO polynomial:
\begin{align}
 y({\bf z}) = & y_0 +\sum_{i_1=1}^{m}\alpha_{i_1}z_{i_1}  +\sum_{i_1=1}^{m}\sum_{i_2=i_1}^{m}\alpha_{i_1,i_2}z_{i_1} z_{i_2} \nonumber \\ 
 +& \sum_{i_1=1}^{m}\cdots \sum_{i_n=i_{n-1}}^{m}\alpha_{i_1,\cdots,i_n}z_{i_1} \cdots z_{i_n}
\end{align}

This coupled polynomial corresponds to the left side of Fig. \ref{fig:decoupled MISO}.  In this section, this will be replaced by the decoupled representation, shown on the right side of the same figure.

As described  above, the MISO polynomial is replaced with a transformation matrix, $\bm {V}$, and several SISO polynomial branches, $g_i(.)$, where $i$ denotes the $i$th branch. Then the outputs of all the branches will be summed together to build the output signal of the MISO system, $y$.

\subsection{Estimation of Decoupled MISO Polynomials} \label{Estimation of D model}
The MISO polynomial will be decoupled using an optimization based approach.  The output of the decoupled model, shown on the right of Figure \ref{fig:decoupled MISO}, can be written: 
\begin{align}
 \label{eq:decoupled}
& \hat y({\bf z})  =   c_0 + \sum_{i = 1}^r g_i({\bm v_i}^T{\bf z}) 
\end{align}
where the summation over $i$ adds the outputs of the $r$ branches together.  Thus, $\bm v_i$ is the $i$th column of the transformation matrix, $\bm V$, and $g_i$ is the univariate polynomial in the $i$th branch, given by: 
\begin{align}
    & g_i(x_i)  =  \sum_{j = 1}^n c_{j,i}x_i^j 
\end{align}
where $x_i= {\bm v_i}^T{\bf z}$ is the input to the polynomial, as shown in (\ref{eq:decoupled}). In this representation, $r$ denotes the number of SISO branches in the decoupled structure and $n$ represents the polynomial degree.

Let ${\bm V} = \left[\begin{array}{cccc} {\bm v_1} & {\bm v_2} & \ldots & {\bm v_r} \end{array}\right]$ be the transformation matrix  and $\mathbf{c}$ be a vector containing all of the polynomial coefficients. Given $N$ data points,
the goal is to find the estimates of $\mathbf{V}$ and $\mathbf{c}$, that minimize the cost function, defined to be the norm of the error between the real output and the estimated output:
\begin{equation}\label{cost}
    V_N(\mathbf{V} , \mathbf{c}) = ||\mathbf{y}-\hat{\mathbf{y}}(\mathbf{V} , \mathbf{c})||_2^2
\end{equation}
where $\mathbf{y}$ and $\hat{\mathbf{y}}$ are $N$ element vectors containing the measured output and estimated output, respectively.

The output of the system can be also expressed as: 
\begin{align} \label{y=xc}
  \hat{\mathbf{y}}= \mathbf{X}\mathbf{c}  
\end{align}
where $\mathbf{X}= \left[\begin{array}{ccccc} {\mathbf{1}_N} & {\mathbf{X}_1} & {\mathbf{X}_2} & \ldots & {\mathbf{X}_r} \end{array}\right]$. ${\mathbf{1}_N}$ is an $N$ element column vector of $1s$ and each $\mathbf{X}_i$ is a Vandermonde matrix based on the signal ${\mathbf{x}_i}$:
\begin{align}
    \mathbf{X_i} & = \left[\begin{array}{cccc} \mathbf{x}_i & \mathbf{x}_i^2 & \ldots & \mathbf{x}_i^n \end{array}\right] \label{vandermonde}\\
   \mathbf{x}_i &= \left[\begin{array}{cccc} x_i(1) & x_i(2) & \ldots & x_i(N) \end{array}\right]^T
\end{align}
and $x_i(t)  =  {\bm v_i}^T{{\bf z}(t)}$ is  the  $i$th intermediate signal evaluated at measurement point $t$.

From Eq.(\ref{y=xc}), $\hat{\mathbf{y}}$ is linear in the polynomial coefficients while the regression matrix, $\mathbf{X}$, depends nonlinearly on the elements of the transformation matrix, $\mathbf{V}$, due to the Vandermonde matrix in Eq.(\ref{vandermonde}).

Thus, the minimization of the cost function can be done using a separable least squares (SLS) optimization \cite{golub2003separable}. The key observation behind SLS optimization is that the parameters which appear linearly, in this case the elements of $\mathbf{c}$, can be written in closed form as functions of the remaining parameters, $\mathbf{V}$. For any given $\mathbf{V}$, the least squares solution for $\mathbf{c}$ is given by:
\begin{align} \label{eq:chat}
   \hat {\mathbf{c}}(\mathbf{V}) = \arg \min_\mathbf{c} || \mathbf{y} - \mathbf{X}\mathbf{c} || 
\end{align}
which can be solved using ordinary least squares methods. Recall that $\mathbf{X}$ is dependent on $\mathbf{V}$ as discussed above. Since $\hat{\mathbf{c}}$ is a function of $\mathbf{V}$ the optimization can be performed with respect to $\mathbf{V}$ only:

\begin{align} \label{eq:vhat}
   \hat{ \mathbf{V}} = \arg \min_{\mathbf{V}} V_N(\mathbf{V},\hat{\mathbf{c}}(\mathbf{V}))
\end{align}
provided that the dependence in Eq.(\ref{eq:chat}) is taken into account.

The optimization can be solved using a quasi-Newton algorithm such as the Gauss-Newton or Levenberg-Marquart algorithm \cite{ljung1998system}. In either case, one needs to compute the Jacobian of the model output. This can be done using the Variable Projection Algorithm \cite{golub2003separable}, which computes a first order approximation to the Jacobian of the separated problem as: 
\begin{align}
    \mathbf{J}_s = \mathbf{J}_v -\mathbf{J}_c(\mathbf{J}_c^T\mathbf{J}_c)^{-1}\mathbf{J}_c^T\mathbf{J}_v
\end{align}
where $\mathbf{J}_v$ and $\mathbf{J}_c$ are the Jacobians of the original, unseparated problem with respect to $\mathbf{V}$ and $\mathbf{c}$ respectively.

The Jacobian of $\hat{\mathbf{y}}$ with respect to the polynomial coefficients is given by: 
\begin{align}
\mathbf{J}_c(k,n*(i-1)+j) &= \frac{\partial \hat{\mathbf{y}}(\mathbf{z}(t))}{\partial c_{i,j}} \nonumber\\
&= ({\bm v}_i^T \mathbf{z}(t))^j 
\end{align}
Similarly, the Jacobian with respect to the elements of mixing vectors are:
\begin{align}\label{eq:dydv}
\mathbf{J}_v(k,m*(i-1)+j) &= \frac{\partial \hat{\mathbf{y}}(\mathbf{z}(t))}{\partial {\bm v}_i(j)} \nonumber \\
&= g_i'({\bm v_i}^T{\bf z}(t))\mathbf{z}(t)(j) 
\end{align}
where $g_i^\prime(x)$ is the derivative, of $g_i$ with respect to its argument.

\subsection{Extension to the NARX Model}\label{sec:extension}

In the case of the polynomial NARX model, which is the model that we are considering in this paper, the input vector $\mathbf{z}$ will depend on the time, $\mathbf{z}(t)$, and will contain past inputs and past outputs to the system which form
the current input to the polynomial.
 \begin{align*}
     \mathbf{z}(t) &= [ y(t-1), \cdots, y(t-n_y),\\
  & \rule{0.5cm}{0cm} u(t-n_k),u(t-n_k-1),\ldots, u(t-n_k-n_u)]
 \end{align*}
where $n_u$ and $n_y$ are the number of past input and output samples used in the model and $n_k$ is the input delay which is always greater than or equal to zero, therefore, the number of polynomial inputs is $m=n_u+n_y$.

Stacking all the measurements of inputs and outputs at $N$ different points together will result in the input matrix $\mathbf{U}$ and the output vector $\mathbf{y}$:
\begin{align*}
     \mathbf{U}& = \left[ \begin{array}{cccc} \mathbf{z}(1) & \mathbf{z}(2) & \ldots & \mathbf{z}(N) \end{array} \right]\\
    \mathbf{y}& = \left[ \begin{array}{cccc} y(1) & y(2) & \ldots & y(N) \end{array} \right]^T
\end{align*}

Having all the polynomial inputs and outputs at different measurement points, $\mathbf{U}$ and $\mathbf{y}$, we can solve the optimization in Eq.(\ref{cost}).

\subsection{Non-Polynomial Nonlinearities}\label{sec:nonpolynomial}
This optimization based approach can be used with any basis expansion model of the nonlinearity (since it will still be linear in the variables), provided that the basis elements are differentiable -- since  the derivative of the nonlinearity is required in Eq.(\ref{eq:dydv}).  For example, spline functions have been routinely used in nonlinear block-structured models such as the Wiener-Hammerstein model \cite{westwick2012initial} and could be used to represent the SISO nonlinearities in the branches.

\section{Optimization Initialization} \label{opt-in}
The method presented in Sec. \ref{Estimation of D model} involves an iterative optimization over a non-convex error surface. As such, a good initial estimate of the optimization variables, $\bm V$, is required. This section will use an approach based on tensor factorization to generate an initial estimate.

\subsection{Factoring of the Jacobian} \label{sec:factoring}
Dreesen et al. \cite{dreesen2015decoupling} proposed a method for decoupling MIMO polynomials that used a tensor constructed by stacking the Jacobian matrices, the partial derivatives of each output with respect to each input,  evaluated at several measurement points.
Stacking these matrices produced a 3-way tensor which was then factored using the Canonical Polyadic Decomposition (CPD) \cite{kolda2009tensor}. 
Kruskal \cite{kruskal1977three} proved the uniqueness of the factors of triple product decomposition.
Although it is a non-convex problem, the CPD can be reliably computed using the Alternating Least Squares (ALS) algorithm.  Wang \cite{wang2014global} has proven the global convergence of the ALS algorithm for rank-one  approximation  to  generic  tensors.

If we follow the approach used in \cite{dreesen2015decoupling} and compute the Jacobian, the result is a vector.
\begin{align} \label{eq:Jy}
\mathbf{J}_y &= \frac{\partial \mathbf{y}(\mathbf{z})}{\partial \mathbf{z}}  
= \sum_{i=1}^{r} g_i'({\bm v_i}^T{\bf z}) {\bm v_i}^T
\end{align}
Stacking all the Jacobians at all $N$ different measurement points forms the matrix
\begin{align} \label{eq:J}
\mathbf{J} = \mathbf{V}\mathbf{W}^T
\end{align}
where $\mathbf{W}$ contains the derivatives of the SISO polynomials with respect to their inputs, evaluated  at all measurement points. 
Since the Jacobian is a matrix, rather than a 3-way tensor, the powerful uniqueness results for tensor factorization no longer apply, since for any invertible transformation matrix (of appropriate size), one can replace $\mathbf{V}\mathbf{W}^T$ with $\mathbf{V} \mathbf{W}^{-1}\mathbf{M}\mathbf{W}^T $(so that the factors become $\mathbf{V}\mathbf{M}^{-1}$ and $\mathbf{M}\mathbf{W}^T$). Thus there are an infinite number of possible factorizations. Given this lack of uniqueness, factoring the Jacobian does not produce a suitable initialization.

\subsection{Factoring of the Hessian} \label{factoring hessian}
In this section, second-order information, namely the Hessian matrix, will be used to generate a tensor which can then be uniquely factorized to reveal the decoupled structure using the CPD.

Evaluating the Hessian of the output of the decoupled model, as in Eq.(\ref{eq:decoupled}), with respect to the inputs (the elements of ${\bf z}$) at the measurement point, $\mathbf{z}(k)$:
\begin{align} \label{eq:hessian}
 \frac{\partial^2 \hat{\mathbf{y}}(\mathbf{z})}{\partial \mathbf{z}^2}  
= \sum_{i=1}^{r} g_i''({\bm v_i}^T{\bf z}){\bm v_i} {\bm v_i}^T \bigg|_{{\bf z}={\bf z}(t)}
\end{align}
If the Hessian is evaluated at $N$ different measurement points, the results may be stacked into a 3-way tensor as: 
\begin{align}\label{eq:Htensor}
\bm{\mathcal {H}}(i,j,k) = \sum_{n=1}^{r} g_n''({\bm v_n}^T{\bf z}(k)){\bm v_n}(i) {\bm v_n}(j)
\end{align} 
This can be written more compactly using the Kronecker product:
\begin{align} \label{eq:H3}
\bm{\mathcal {H}} = \sum_{n=1}^{r}{\mathbf{w}_n} \otimes {\bm v_n} \otimes {\bm v_n}
\end{align}
where the vector $\mathbf{w_n} \in \Re^N$ can be defined for $n=1,\cdots,r$, to contain the second derivative of the polynomial in branch $n$ with respect to its input, evaluated at each of the measurement points:
\begin{align} \label{eq:wn}
    \mathbf{w}_n  = \left[\begin{array}{cccc} g_n''({\bm v_n}^T{\bf z}(1)) & \ldots &  g_n''({\bm v_n}^T{\bf z}(N)) \end{array}\right]^T 
\end{align}
The decomposition in Eq.(\ref{eq:H3}) is the CPD, the same decomposition used by Dreesen et al. \cite{dreesen2015decoupling} to decouple MIMO polynomials,
and can be computed using standard tools such as those in Tensorlab \cite{vervliet2016tensorlab}. 

One of the major difficulties with factoring the Hessian in Eq.(\ref{eq:H3}) is that the number of variables in $\mathbf{W}$ increases in proportion with the number of measurement points $N$  since in Eq.(\ref{eq:wn}) $\mathbf{w}_n \in \Re^{N\times 1}$.  Thus, unlike the $\mathbf{V}$ factor, whose dimensions are independent of the number of measurements, increasing N will not reduce the sensitivity of the $\mathbf{W}$ estimate to noise.  One solution to this problem, is to impose structure on the $\mathbf{W}$ factor, hence reducing the number of its parameters.  
 
As shown in Fig. \ref{fig:initialCPD-SB} and Fig. \ref{fig:initialCPD}, when no structure is added during the CPD of a noisy tensor, the decomposed factors are noisy. In particular, clouds of points will appear when plotting the columns of $\mathbf{w}$ against ${\bm v}^T{\bf z}$. Specifically, these clouds of points in the factor $\mathbf{W}$ can hardly represent the second derivatives of the decoupled functions. Note that similar  problems occur in the CPD of the Jacobian tensor used to decouple MIMO polynomials \cite{gabriel-phd}. Hollander addressed this by imposing a polynomial structure on the elements of the CPD factor that represents the polynomial derivatives, $\mathbf{W}$.

\subsubsection{Imposing the Polynomial Structure}\label{sec:polynomial}
Instead of solving directly for the elements of $\mathbf{W}$, we will solve for the coefficients of the polynomials $g_n''$ in Eq.(\ref{eq:Htensor}).
By adding this polynomial constraint during the CPD, the output factor $\mathbf{W}$ now represents directly the second derivatives of the functions $g_1,\cdots,g_r$. 
In order to impose the polynomial structure we will update the coefficients of the second derivatives of $g_1,\cdots,g_r$ instead of directly updating the entries in $\mathbf{W}$.

We will solve an optimization problem that minimizes the mean squared error between the Hessian and its approximation. Since a quasi-Newton optimization will be used to update the polynomial coefficients, we will calculate the Jacobian of the Hessian, the partial derivatives of the Hessian with respect to the mixing matrix elements, $\mathbf{V}$, and the polynomial coefficients, $\mathbf{C}$,  as shown in Eqs.(\ref{eq:dHdv}) and (\ref{eq:dHdc}): 
\begin{align}\label{eq:dHdv}
    \frac{\partial \mathbf{H}(i,j,k)}{\partial  \mathbf{V}(m,l)} & =  \frac{\partial}{\partial \mathbf{V}(m,l)} \sum_{n=1}^{r} \mathbf{V}(i,n)\mathbf{V}(j,n)\mathbf{W}(k,n) \\ \nonumber
    & = \frac{\partial}{\partial \mathbf{V}(m,l)}  \big( \mathbf{V}(i,l)\mathbf{V}(j,l)\mathbf{W}(k,l)\big) \\ \nonumber
    & =  \frac{\partial \mathbf{V}(i,l)}{\partial \mathbf{V}(m,l)} \mathbf{V}(j,l)\mathbf{W}(k,l)\\ \nonumber
    & + \mathbf{V}(i,l) \frac{\partial \mathbf{V}(j,l)}{\partial \mathbf{V}(m,l)} \mathbf{W}(k,l) \\ \nonumber
    & + \mathbf{V}(i,l) \mathbf{V}(j,l) \frac{\partial \mathbf{W}(k,l)}{\partial \mathbf{V}(m,l)}
\end{align}
where $i,j, m = 1,\cdots,(n_a+n_b)$, $k=1,\cdots,N$ and $l=1,\cdots,r$.
Simplifying Eq. (\ref{eq:dHdv}): 
\begin{align}
    \frac{\partial \mathbf{H}(i,j,k)} {\partial \mathbf{V}(m,l)} & = 
    \delta_{im}\mathbf{V}(j,l)\mathbf{W}(k,l)\\ \nonumber
    & + \delta_{jm}\mathbf{V}(i,l)\mathbf{W}(k,l)\\ \nonumber
    & + \mathbf{V}(i,l)\mathbf{V}(j,l)\frac{\partial \mathbf{W}(k,l)}{\partial \mathbf{V}(k,l)}\\ \nonumber
\end{align}
where $\delta$ is a the Kronecker delta function and
\begin{align} \label{eq:W}
    \mathbf{W}(k,l) = \frac{\partial^2 g_i(x_i)}{\partial x_i^2} = \sum_{j = 1}^{M} j(j-1)c_{j,i}x_i^{(j-2)} 
\end{align}

\begin{align} \label{eq:dHdc}
    \frac{\partial \mathbf{H}(i,j,k)}{\partial \mathbf{C}(t,l)} & =  \frac{\partial}{\partial \mathbf{C}(t,l)} \sum_{n=1}^{r} \mathbf{V}(i,n) \mathbf{V}(j,n)\mathbf{W}(k,n) \\ \nonumber
    & =  \mathbf{V}(i,l) \mathbf{V}(j,l) \frac{\partial \mathbf{W}(k,l)}{\partial \mathbf{C}(t,l)}
\end{align}
Note that storage requirements for Jacobian of the Hessian are large -- the Hessian contains $Nm^2$ elements, and the model contains $r(m+M-2)$ parameters.
Thus the Jacobian will have $Nrm^2(m+M-2)$ elements. The size of this matrix scales approximately with $m^3$, the cube of the number of inputs in the vector ${\bf z}$. 

We will minimize the norm of the error between the Hessian and its decoupled approximation using Eqs.(\ref{eq:H3}) and (\ref{eq:W}). The tensor values are nonlinear in the mixing matrix ($\mathbf{V}$) elements but linear in the polynomial coefficients ($\mathbf{C}$). Therefore, we will treat $\mathbf{C}$ as a function of $\mathbf{V}$ and optimize over the mixing matrix. This optimization decomposes the tensor, but with the addition of the polynomial constraint.

\begin{figure}[h!]
\centering
\includegraphics[scale=0.5]{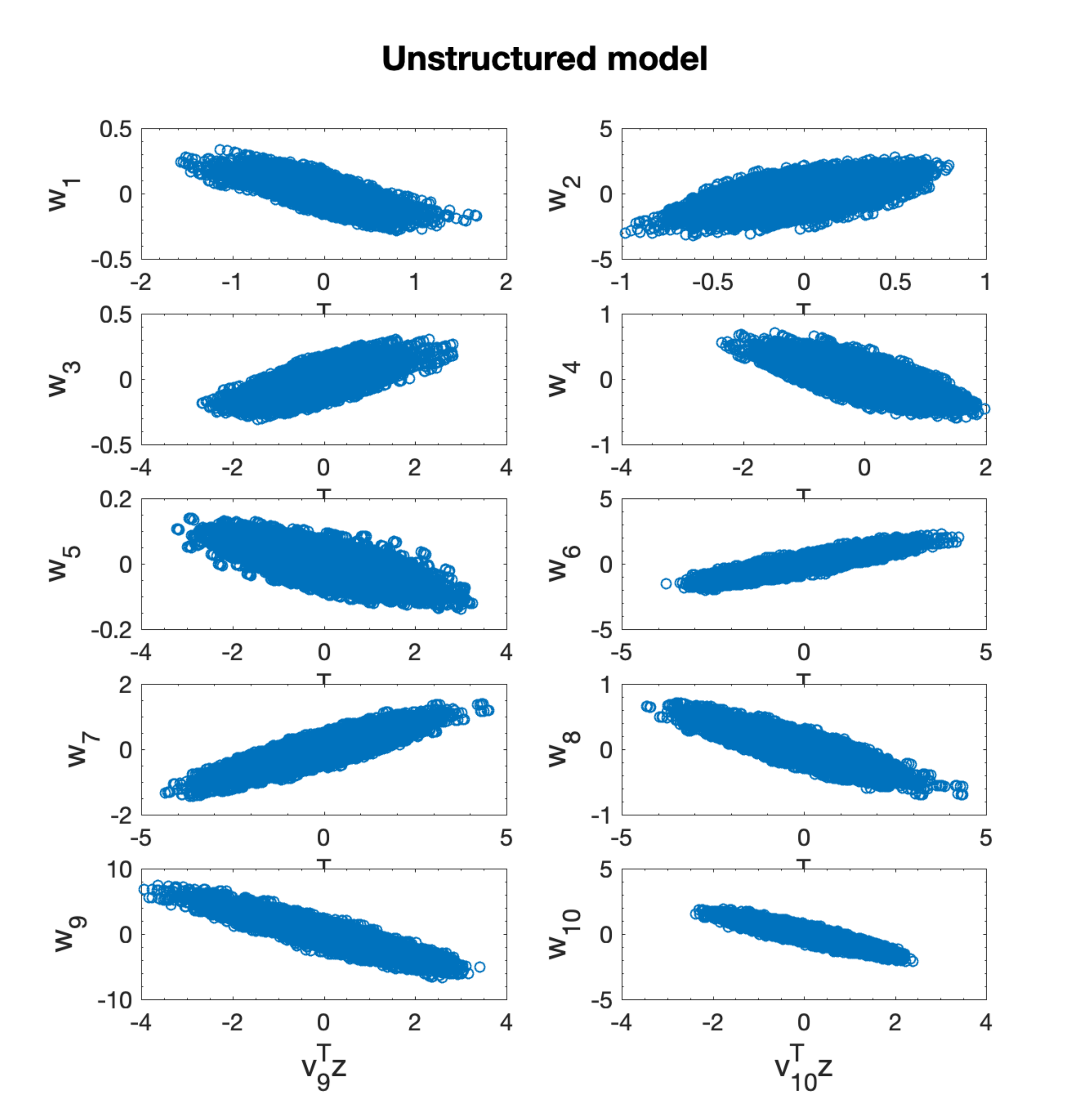}
\caption{Visualization of the Vectors $\mathbf{w}$ Which Contain the Estimates of the Mapping Functions Obtained From the CPD of the Hessian Using Bouc-Wen Data}
\label{fig:initialCPD}
\end{figure}

\begin{figure}[h!]
\centering
\includegraphics[scale=0.5]{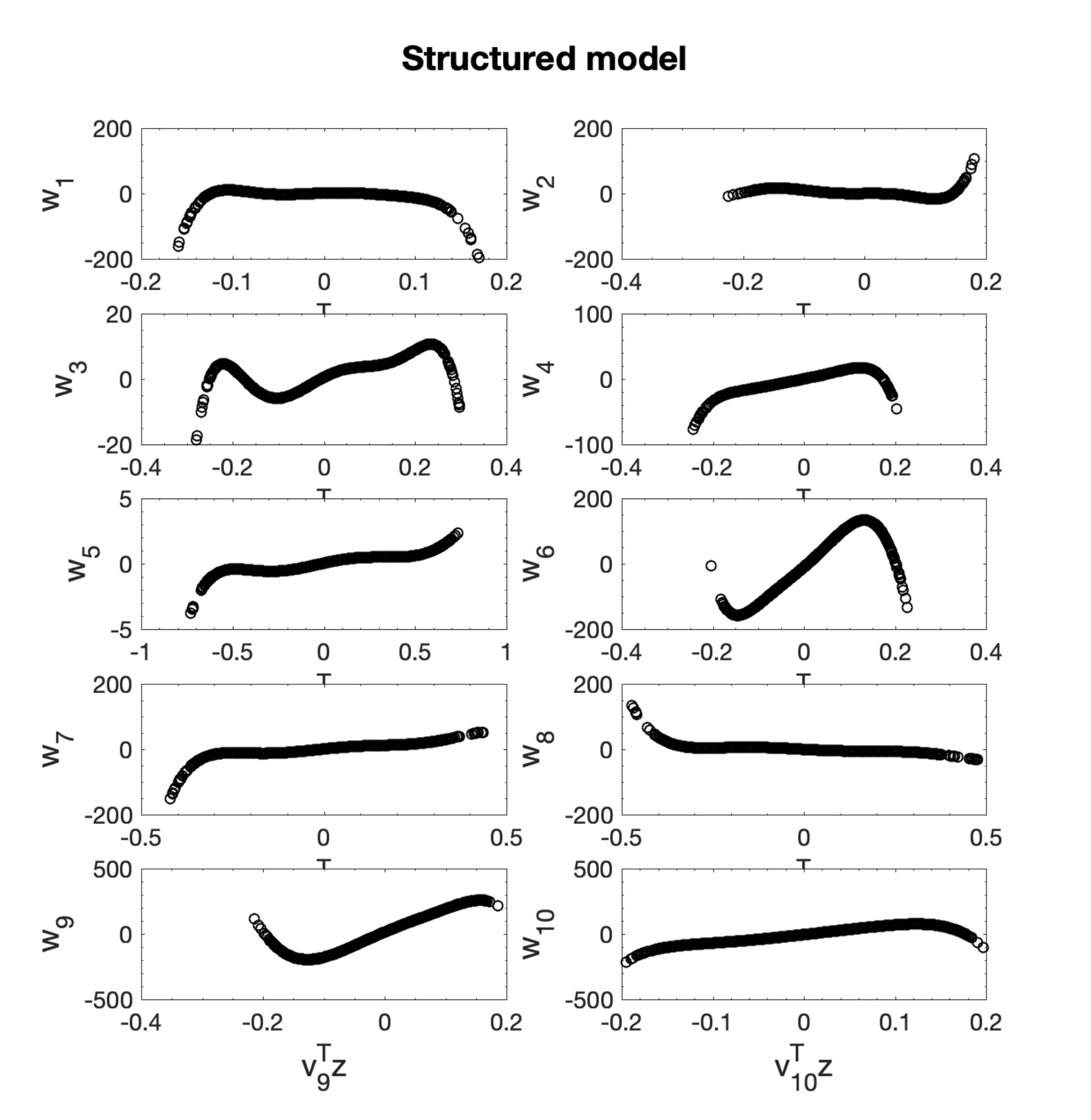}
\caption{Visualization of the Vectors $\mathbf{w}$ Which Contain the Estimates of the Mapping Functions Obtained From Polynomial structure enforced on the  CPD of the Hessian Using the Same Data as in Figure \ref{fig:initialCPD}}
\label{fig:Initial CPDPoly}
\end{figure}

In Figure \ref{fig:Initial CPDPoly} the polynomial structure is imposed. The plots for Fig. \ref{fig:Initial CPDPoly} are generated from the same experiments as in Fig. \ref{fig:initialCPD}. The performance of the identified model using different initial solution will be examined in Section \ref{benchmark}.

\subsection{Use of Non-polynomial Bases} \label{sec:nonPbasis}
The Hessian based decoupling can be applied in any case where the multivariate basis elements are the products of single-variable terms.
The derivatives of this structure, and hence the resulting Jacobian matrix and Hessian tensor, will share the same structure, as in the polynomial case, with one factor, $\mathbf{V}$, that contains the mixing matrix elements, and another factor, $\mathbf{W}$, that is contains the first or second derivatives of the nonlinearities. Since we really only use the $\mathbf{V}$ matrix in the decoupling, and all of the complication arising from the alternate basis elements will turn up in the $\mathbf{W}$ factor, the decoupling algorithm can be applied with little modification.

\section{Algorithm} \label{alg}
The following algorithm identifies a decoupled polynomial NARX (DP-NARX) model from input-output data. Note that the algorithm assumes that the user has chosen the number of past inputs and outputs and the input delay:

\begin{algorithm}[H] 
\caption{Decoupling a Polynomial NARX Model from Input-Output Data Initialized by Decomposing the Hessian using a Polynomial Constraint}
1. Fit a polynomial NARX model to the provided data using standard approaches such as an orthogonal forward regression algorithm \cite{billings2013nonlinear} \\ [0.5ex]
2. Compute the Hessian with respect to the inputs and outputs  using Eq.(\ref{eq:hessian})\\ [0.5ex]
3. Impose the polynomial structure on the Hessian as described in Section \ref{sec:polynomial}\\ [0.5ex]
4. Compute the CPD of the unconstrained Hessian using standard tools \cite{vervliet2016tensorlab} \\ [0.5ex]
5. Use the  $\mathbf{V}$ from the CPD  as an initial estimate of the mixing matrix in the decoupled model. Refine the initial estimate using a SLS based optimization Eq.(\ref{eq:vhat})
\label{alg:alg3}
\end{algorithm}

\section{Benchmark Results} \label{benchmark}
The polynomial NARX decoupling algorithm is applied to 2 benchmark data sets: measurement data from an electronic implementation of a forced Duffing Oscillator  \cite{wigren2013three} and simulation data from a nonlinear differential equation based friction model \cite{noel2016hysteretic,schoukens2017three}.

\subsection{Application to the Silver-Box Benchmark}
\begin{figure}[h]
\begin{center}
\includegraphics[scale = 0.8]{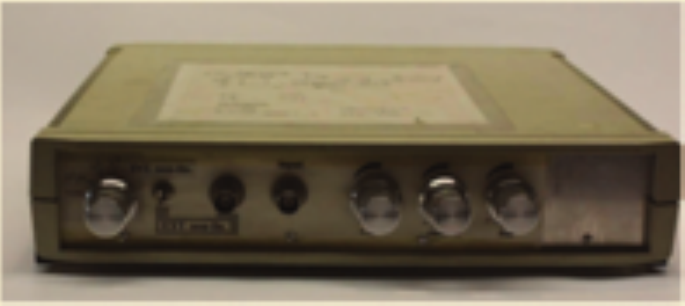}\\
\includegraphics[scale = 0.8]{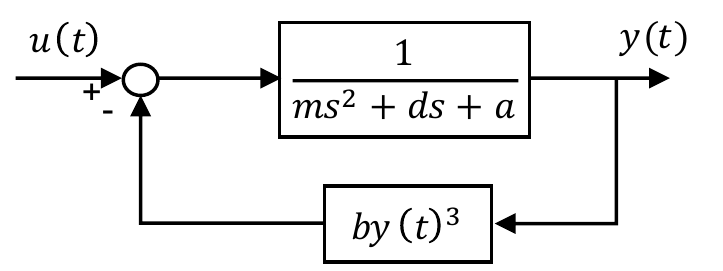}  
\caption{Upper: Real Silver-Box, Lower: Schematic Representation of Silver-Box \cite{schoukens2016linear}} 
\label{fig:Silverbox}
\end{center}
\end{figure}

The Silver-Box \cite{wigren2013three} is an electronic implementation of a forced Duffing oscillator: a nonlinear mechanical resonating system incorporating a moving mass \textit{m}, a viscous damping \textit{d}, and a nonlinear spring \textit{k(y)}. As shown in Fig. \ref{fig:Silverbox}, this implementation is a second order closed-loop system with a simple nonlinearity in the feedback path.
The electrical circuit represents the relationship between the force, which is the input voltage \textit{u(t)}, and the displacement, represented by the output voltage \textit{y(t)}, using the differential equation: 
\begin{align} \label{eq:silverbox1}
m{{d^2 y(t)}\over{dt}}+d {{d y(t)}\over{dt}}+k(y(t))y(t) = u(t)
\end{align}
The nonlinear spring is described by Eq.(\ref{eq:silverbox2}) which is  a static but position-dependent stiffness: 
\begin{align} \label{eq:silverbox2}
k(y(t)) = a+by^2(t)
\end{align}

\begin{figure}[h]
\centering
\includegraphics[scale=0.35]{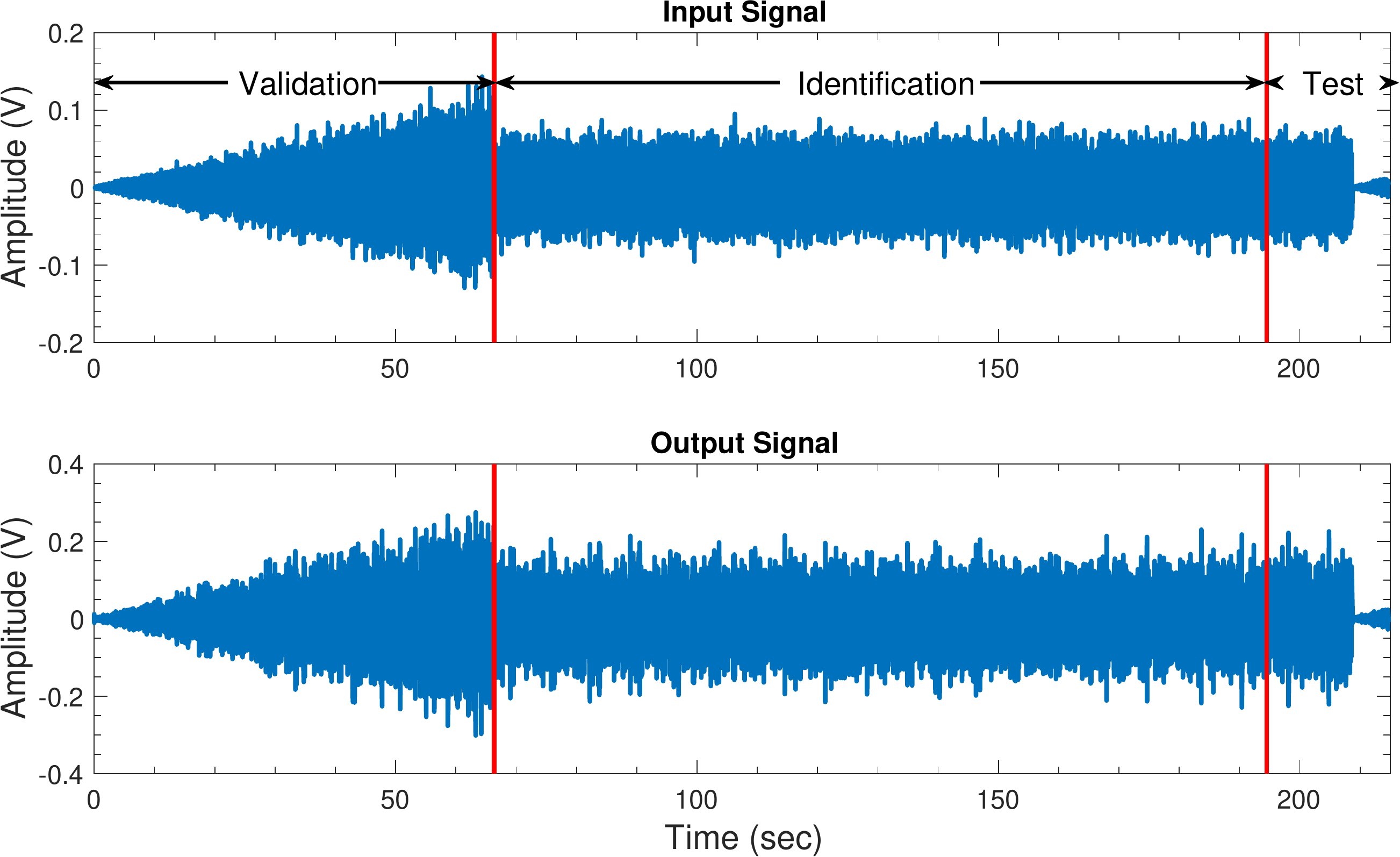}
\caption{Silver-Box Reference Signal. Upper: Input Signal. Lower: Output Signal}
\label{fig:silverboxdata}
\end{figure}

The output signal to noise ratio is estimated to be about $40$dB, which is high enough to warrant modeling without measurement noise.
The data sets for this benchmark can be downloaded using the links provided in (\url{www.nonlinearbenchmark.org}). As Fig. \ref{fig:silverboxdata} illustrates, the reference signal consists of two parts: the first part ($40000$ samples) which is used for validation of the obtained model is a white Gaussian noise sequence filtered by a $9^{th}$ order discrete time Butterworth filter with a $200$Hz cut-off frequency. 
The next part of the reference signal, which is used for identification and testing purposes, consists of $10$ realizations ($9$ realizations used for identification and $1$ realization for testing) of a random phase odd multi-sine. The period of the multi-sine is $8192$ samples.  The data were sampled at $f_s = 610.35$Hz. Each realization of the random odd multi-sine is followed by $100$ zeros (as an indicator to show the start of a new realization). 
Model accuracy is calculated using the RMS model fit:
\begin{align} \label{eq:fit}
    FIT = 100\times \bigg(1-\sqrt{\frac{\sum_{t=1}^{N_t}(y(t)-\hat{y}(t))^2}{\sum_{t=1}^{N_t}(y(t)-y_{avg})^2}}\bigg)
\end{align}

The maximum number of parameters in a polynomial NARX model depends on the number of past inputs and outputs included in the model ($m = n_u+n_y$), and on the order of the polynomial. For a $3rd$ degree polynomial nonlinearity, the maximum number of parameters is:

 \begin{align} \label{eq:narx-param}
    P_{PNARX} = 1 + m + \frac{{m}({m}+1)}{2} + \frac{m({m}+1)({m}+2)}{6}
\end{align}
Note that the parameter values are typically estimated using an orthogonal forward regression \cite{billings2013nonlinear} that can eliminate a large number of insignificant terms from the model.
On the other hand, in case of a decoupled polynomial NARX model, the number of parameters is equal to 
\begin{align}
    P_{DPNARX}= (({m}+M)\times r)+1
\end{align}
where $M$ is the nonlinearity degree and $r$ is the number of branches in decoupled model.

For the Silver-Box benchmark, Eq.(\ref{eq:narx-param}) gives $84$ parameters, but the orthogonal forward regression algorithm eliminated $24$ parameters so the full P-NARX model actually had $60$ parameter versus $37$ parameters in the final decoupled polynomial NARX model. Even for a model as small as the Silver-Box (only 3 past inputs and 3 past outputs), there is a significant reduction in the number of parameters.

\begin{figure}[h]
\centering
\includegraphics[scale = 0.4]{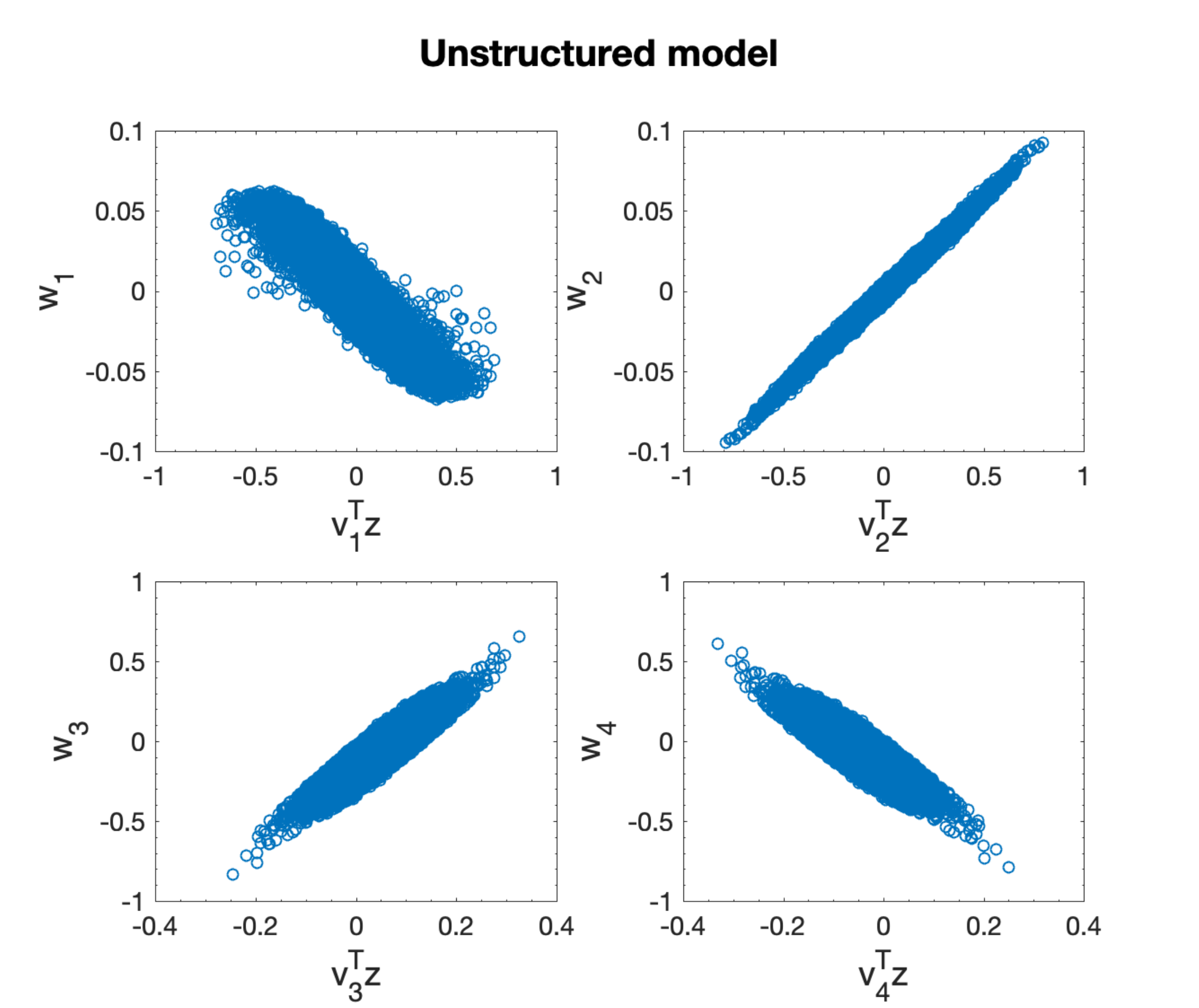}
\caption{Graphics of $\mathbf{w}$ Vectors Containing the Estimates of the Mapping Functions Component Obtained From the CPD of the Hessian Using Silver-Box Data }
\label{fig:initialCPD-SB}
\end{figure}

\begin{figure}[h]
\centering
\includegraphics[scale = 0.4]{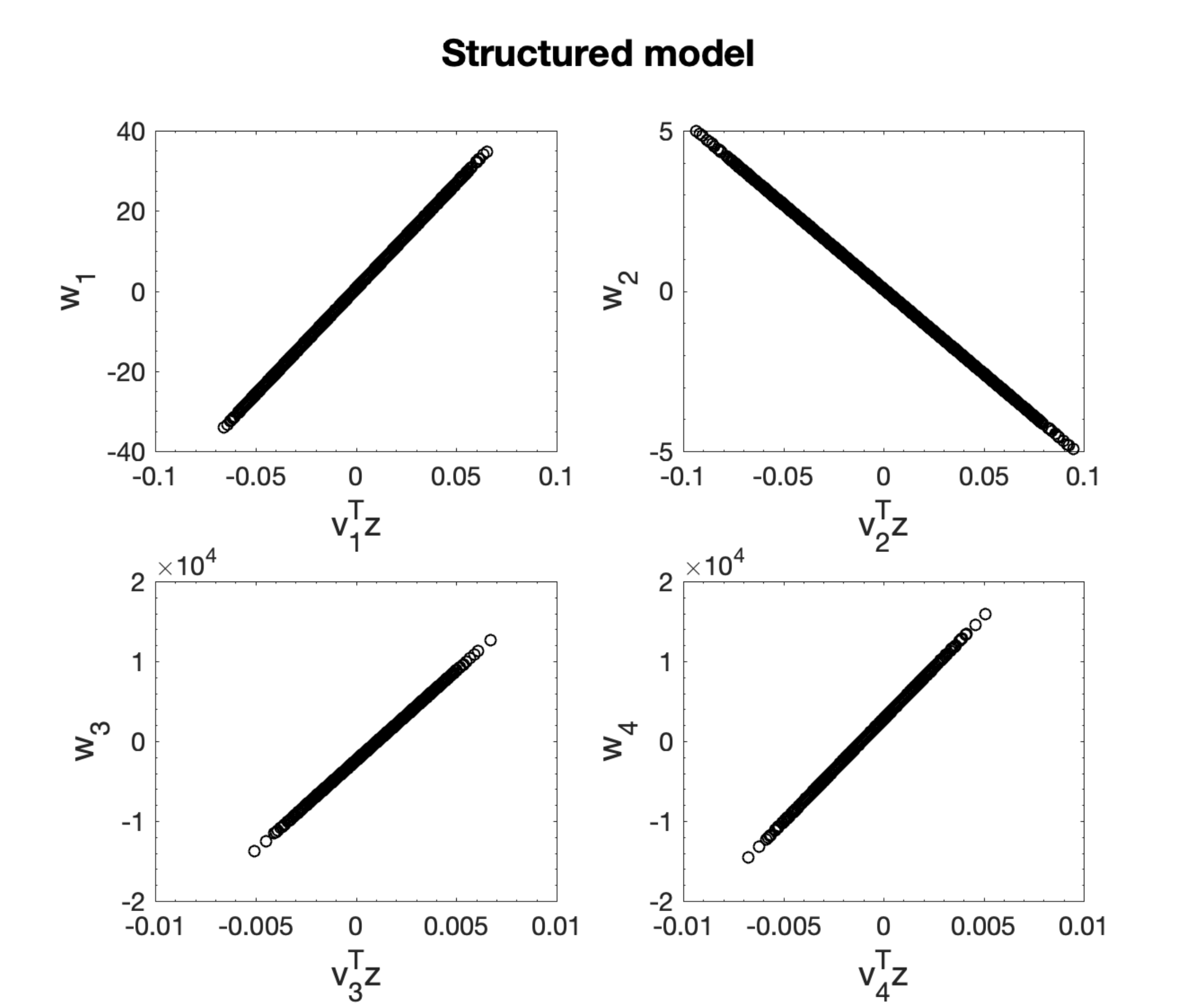}
\caption{Graphics of $\mathbf{w}$ Vectors Containing the Estimates of the Mapping Functions Component Obtained From Polynomial Structure Enforced on the CPD of the Hessian Using the Same Data as in Figure \ref{fig:initialCPD-SB}}
\label{fig:initialCPDPoly-SB}
\end{figure}

Figures \ref{fig:initialCPD-SB} and \ref{fig:initialCPDPoly-SB} show 
the CPD factors that represent the second derivatives of the polynomials -- with the constraint, these are straight lines, without it they are point clouds.

\begin{figure}[h]
\centering
\includegraphics[scale = 0.5]{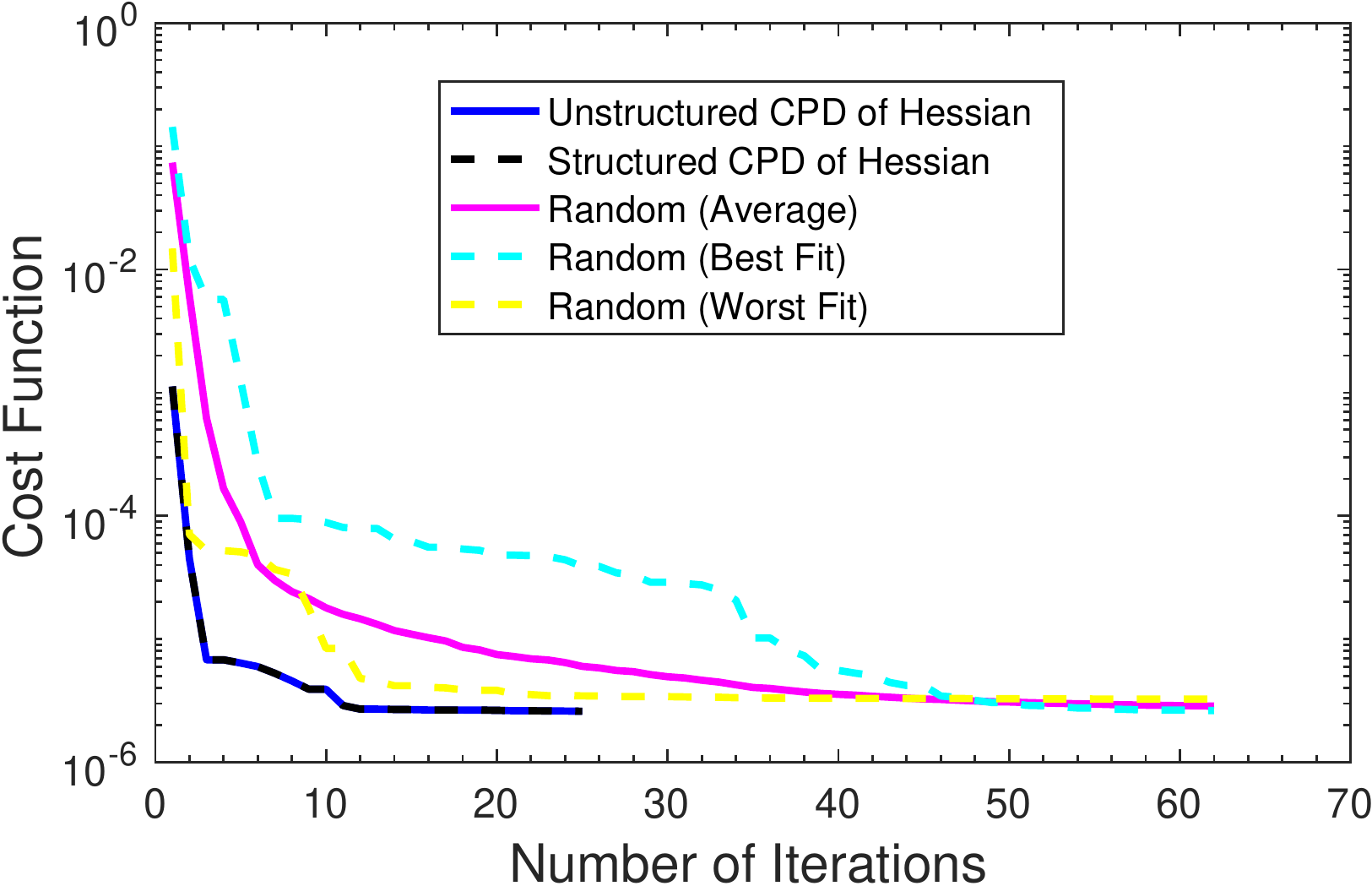}
\caption{Cost Function During Optimization of Decoupled Polynomial NARX Models of the Silver-Box Benchmark}
\label{fig:Cost-SB}
\end{figure}

Figure \ref{fig:Cost-SB} illustrates the cost function with respect to the model output, in the final optimization , where optimization starts from different initial values: 100 random initialization CPD of the Hessian without any structure enforcement and CPD with polynomial structure imposed. As this figure shows, with random initialization it takes $2$-$3$ times as many iterations for the cost function to converge, and never resulted in better models than either of the CPD-based initialization methods. 
As Table \ref{tbl:SBresults} confirms, both CPDs also provide better fit than the random initialization on the validation data.

\begin{table}[h]
\centering
\begin{tabular}{c|c|c}
Initial Solution & Fit\%(Test) & Fit\%(Triangle) \\
\hline
Unstructured CPD of Hessian & 99.82\% & 99.77\% \\
Structured CPD of Hessian & 99.82\% & 99.77\% \\
Random (Average) & 99.80\% & 99.74\% \\ 
\end{tabular}
\caption{Accuracy in the Validation Data Set of 1-Step Ahead Prediction for the Silver-Box Model}
\label{tbl:SBresults}
\end{table}

\begin{figure}[h]
\centering
\includegraphics[scale = 0.4]{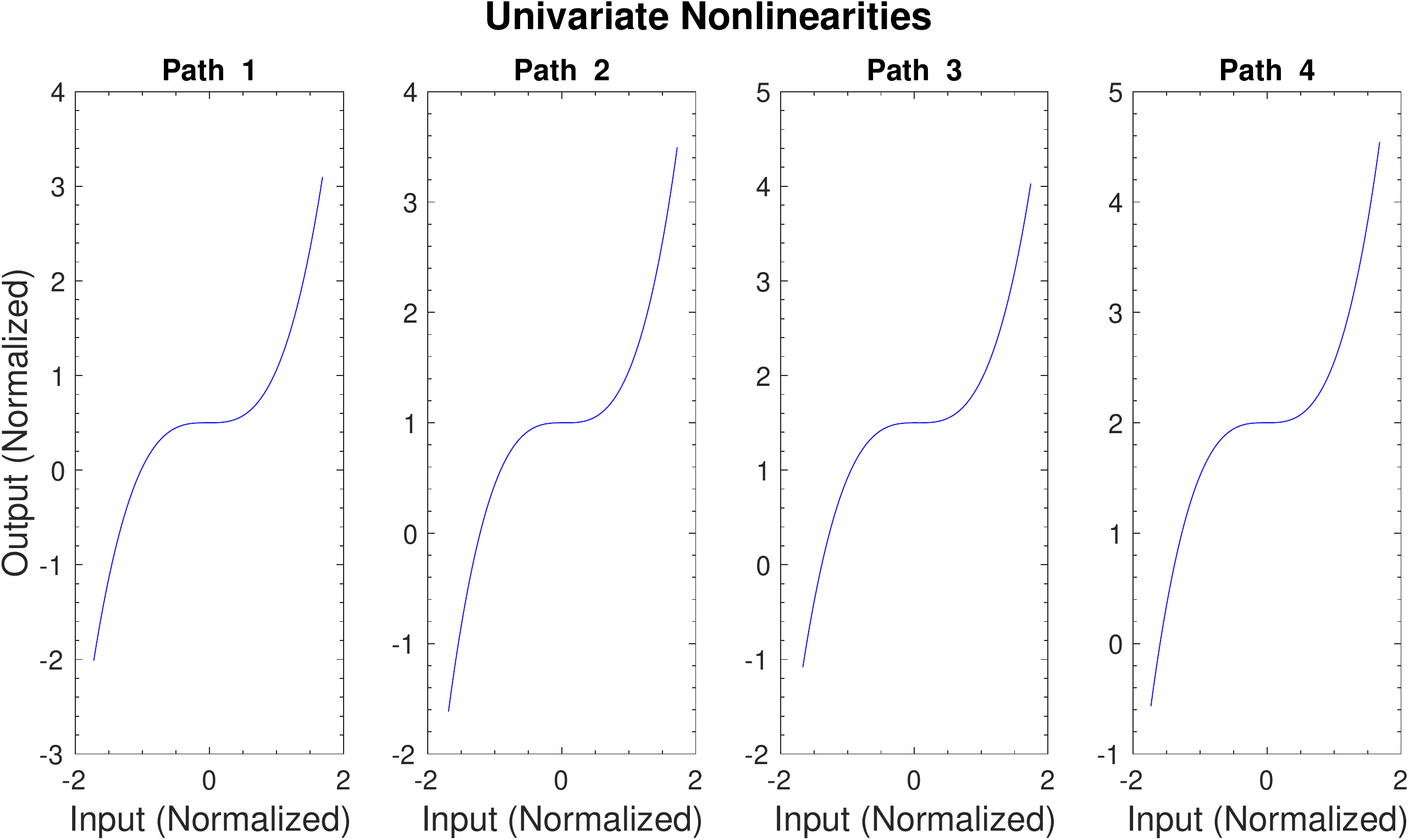}
\caption{Nonlinearities of All 4 Branches of the   Silver-Box Model. The Linear Terms  Have Been Removed to Emphasize the Nonlinear Aspects of the Functions}
\label{fig:unipoynomialCPD-SB}
\end{figure}

\begin{figure}[h]
\centering
\includegraphics[scale=0.5]{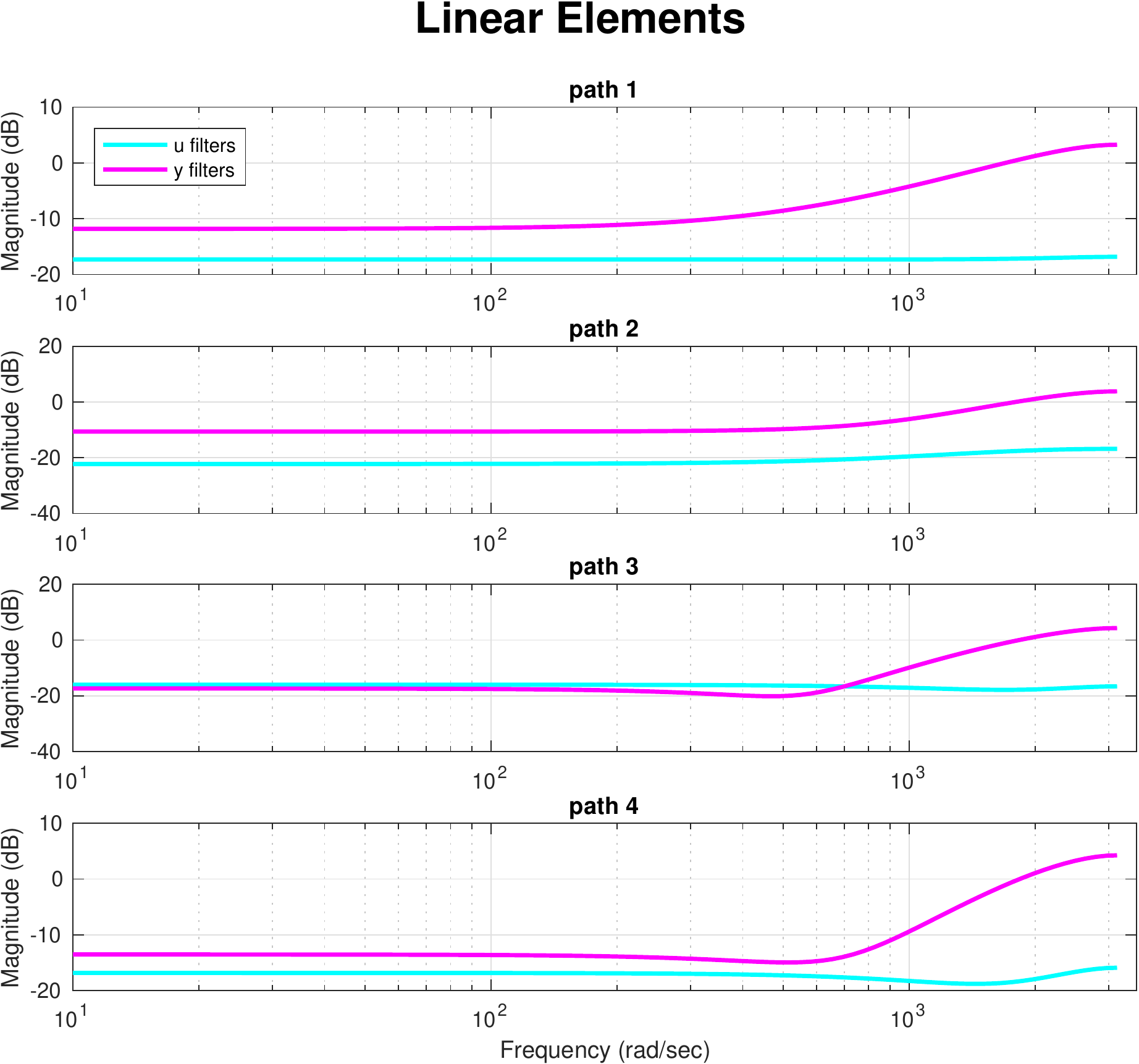}
\caption{Frequency Response Magnitudes for the Input Terms (Cyan) and Output Terms (Magenta) of the Filters in Each Branch of the Silver-Box Model}
\label{fig:uyfilterCPD-SB}
\end{figure}

Figures \ref{fig:unipoynomialCPD-SB} and \ref{fig:uyfilterCPD-SB} show the components of the individual branches of the decoupled model. There are $4$ branches and each branch contains a $3^{rd}$ degree nonlinearity. Fig. \ref{fig:unipoynomialCPD-SB} shows the nonlinearity of each branch. The inputs and outputs of nonlinearites have been normalized.

Note that the filter in each branch receives the output of the entire system, thus, it is not appropriate to combine the input and output terms into a single transfer function.
Thus, the linear element in each branch is treated as two finite impulse response filters: one applied to the input signal, while the other is applied to the overall system output $(y(t))$.
Figure \ref{fig:uyfilterCPD-SB} shows the magnitude frequency responses of these filters. The output filters (shown with magenta lines) are all high-pass in nature.

\begin{figure}[h]
\centering
\includegraphics[width = 85mm]{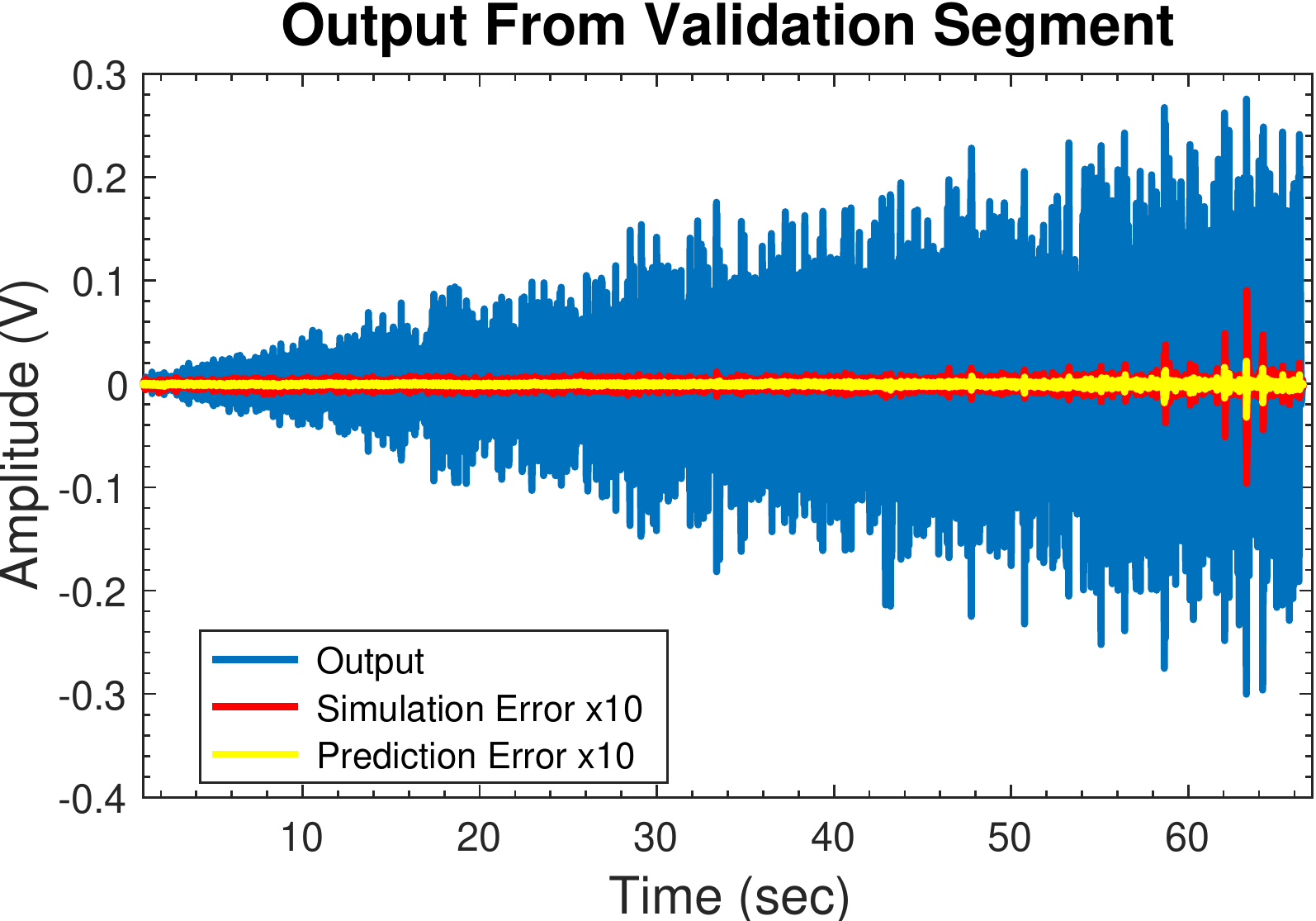}
\caption{Prediction and Simulation Errors on the Validation Data for Silver-Box. Errors Are Scaled Up 10x for Better Visibility}
\label{fig:triangle-SB}
\end{figure}

Figure \ref{fig:triangle-SB} and Table \ref{tbl:SBresults2} show the accuracy of identified model in the triangular validation data. Both 1-step ahead prediction error and simulation error are shown. In the prediction error, shown in yellow, the model has access to past inputs and outputs whereas in the simulation error, shown in red, the model only has access to past outputs. Comparing the results in Table \ref{tbl:SBresults2} to what we obtained previously (simulation accuracy $=98.81\%$) in \cite{westwick2018using} shows higher simulation accuracy using a better initial solution for optimization.

\begin{table}[H]
\centering
\begin{tabular}{c|cc|cc}
   & \multicolumn{2}{c|}{ Hessian Initialization} & \multicolumn{2}{c}{Random Initialization} \\ 
   &  Fit\%  & $e_{RMS}$ & Fit\% & $e_{RMS}$ \\
\hline
Prediction & 99.77\% &0.00012 &99.76\% & 0.00012 \\
Simulation & 99.11\% & 0.00047 &98.81\% & 0.00054 \\ 
\end{tabular}
\caption{Validation Results for Decoupled P-NARX Models on the Arrowhead Data.  Column 1 Shows the Results Obtained with the Model Initialized Using the CPD of the Hessian Tensor. Column 2, Also Shown in \cite{westwick2018using} Reports the Result from the Best Model Obtained Using a Set of 100 Random Initializations.}
\label{tbl:SBresults2}
\end{table}

\subsection{Application to the Bouc-Wen Benchmark} \label{boucwen}
In this section, the decoupled polynomial NARX model is fitted to data from the Bouc-Wen benchmark system, described in detail in \cite{noel2016hysteretic}. Concisely, this benchmark system consists of a single mass, $m_L$, spring, $k_L$, and damper system, $c_L$, with a hysteretic friction term, $z(.)$. In continuous-time, the system is governed by the second-order differential
equation:
\begin{align} \label{eq:bouc1}
m_L\Ddot{y}(t)+c_L\Dot{y}(t)+k_Ly(t) +z(y(t),\Dot{y}(t)) = u(t)  
\end{align}
where $y(t)$ is the displacement and $u(t)$ represents the external force. The hysteretic friction term, obeys this differential equation:
\begin{align}\label{eq:bouc2}
\Dot{z}(y(t),\Dot{y}(t))& = \alpha\Dot{y}(t) \nonumber \\
& -\beta(\gamma|\Dot{y}(t)||z(t)|^{\nu-1}z(t)+\delta\Dot{y}(t)|z(t)|^\nu)
\end{align}

where the values of these parameters are given in Table \ref{tbl:BWparam}.
\begin{table}[H]
\centering
\begin{tabular}{c|cccccccc}
Parameter & $m_L$ & $c_L$ & $k_L$ & $\alpha$ & $\beta$ & $\gamma$ & $\delta$ & $nu$\\
\hline
Value (SI unit) & 2 & 10 & $5\times10^4$ & $5\times10^4$ & $1\times10^3$ & 0.8 & -1.1 & 1
\end{tabular}
\caption{Physical Parameters of the Bouc-Wen System}
\label{tbl:BWparam}
\end{table}

\begin{figure}[h!]
\centering
\includegraphics[width = 85mm,height = 85mm]{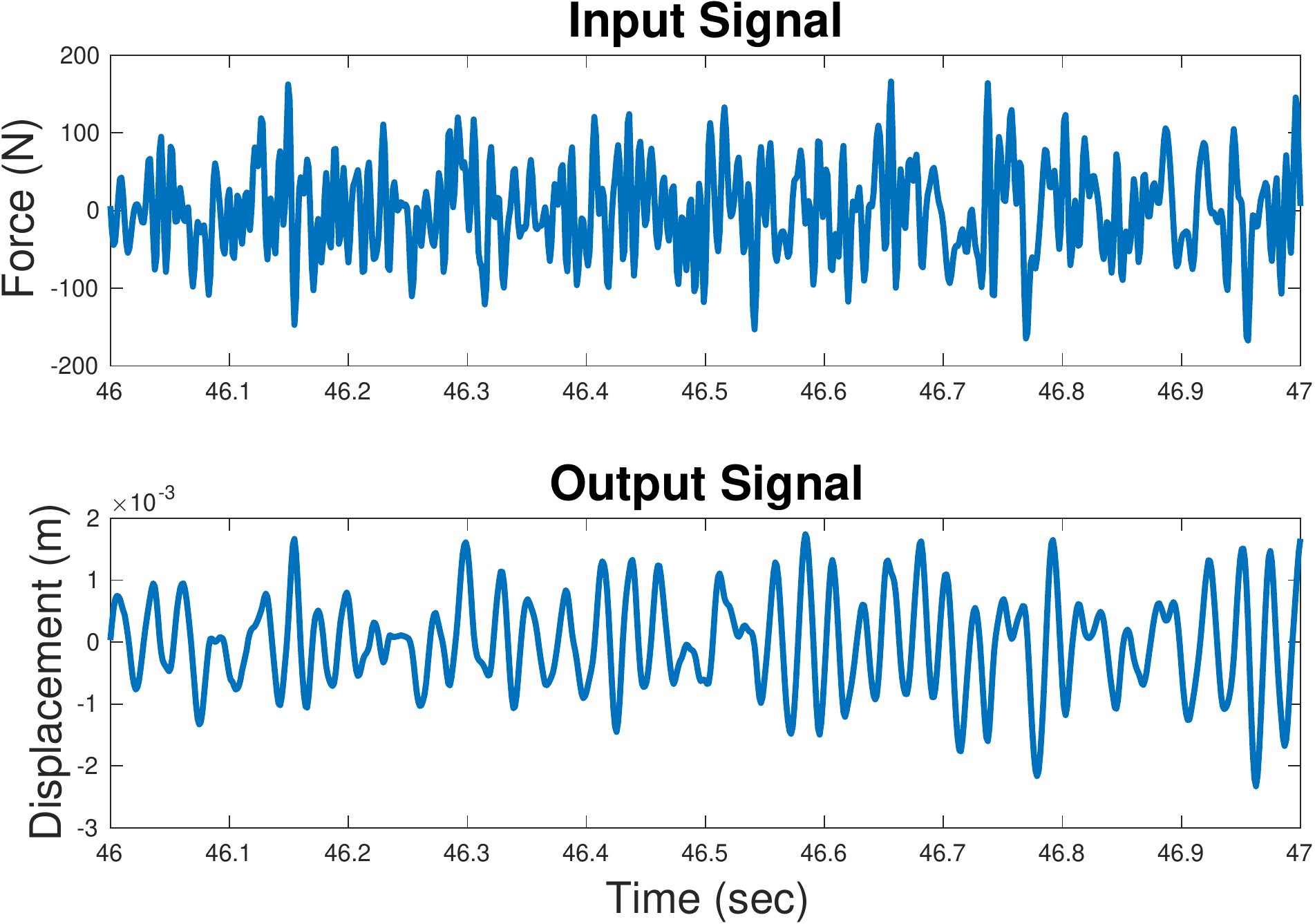}
\caption{ Part of the Identification Data from the Bouc-Wen System. Upper: Input Signal. Lower: Output Signal}
\label{fig:BWdata}
\end{figure}

By integrating Eqs.(\ref{eq:bouc1}) and (\ref{eq:bouc2}) and using the Newmark method, simulation data for Bouc-Wen system is generated. Figure \ref{fig:BWdata} shows the data for the input and the output of Bouc-Wen benchmark. Two data-sets were generated: one was used for parameter estimation, while the second was used for structure selection. These data are generated using the default multi-sine input excitation but, as suggested in \cite{esfahani2018parameter}, the RMS force was increased from $50$N to $55$N, hence, the identification data would cover a larger range than the validation sets. Also, two validation data sets are provided by the benchmark: a multi-sine input and a swept-sine input. The performance of the model is presented using the RMS error in Eq.(\ref{RMS}) as recommended by the benchmark:
\begin{align}
\label{RMS}
    e_{RMS}= \sqrt{\frac{1}{N} \sum_{t=1}^{N}(y_{model}(t)-y(t))^2}
\end{align}

One of the challenges with this system is that the nonlinearity is applied to an internal state $z(.)$ which can not be measured directly, whereas the model that we are using in this work, the polynomial NARX model, is an input-output model. So in order to model the system's nonlinear structure, the internal state has to be reconstructed from just the input and output measurements.

The decoupled polynomial model used in this work contains $4$ past inputs and $6$ past outputs with no input delay. There are $10$ branches in the chosen model each of which contains an $8^{th}$ degree polynomial nonlinearity. This model is smaller than what we had chosen as the best model for this system in previous works \cite{westwick2018using,karami2019identification}. The reason is that, as described in Section \ref{sec:polynomial}, the storage required for the Jacobian of the Hessian scales with the cube of the number of inputs.
Table \ref{tab:numberofJacElement} shows the number of elements in the Jacobian for both models, the one in \cite{westwick2018using} and the reduced model in this paper, where indicating a $10\times$ reduction in the required storage.
As a result, it was infeasible to factor the Hessian tensors of those larger structures.  Thus,  a smaller model with acceptable performance was chosen.
\begin{table}[]
    \centering
    \begin{tabular}{c|ccc|c|c}
         & M & r & m & \makecell{\# of Elements \\ in Jacobian} & \makecell{Required \\ Memory} \\ 
         \hline
         Model in \cite{westwick2018using} & 9&5&30&6.8198$\times10^9$ & 50.8 GB\\
         \hline
         \makecell{Reduced Model in \\ This Paper}&8&10&10&6.5536$\times10^8$ & 4.9 GB
    \end{tabular}
    \caption{The Number of Elements in the Jacobian and the Required Memory for Two Different Models}
    \label{tab:numberofJacElement}
\end{table}

\begin{figure}[h!]
\centering
\includegraphics[width = 85mm]{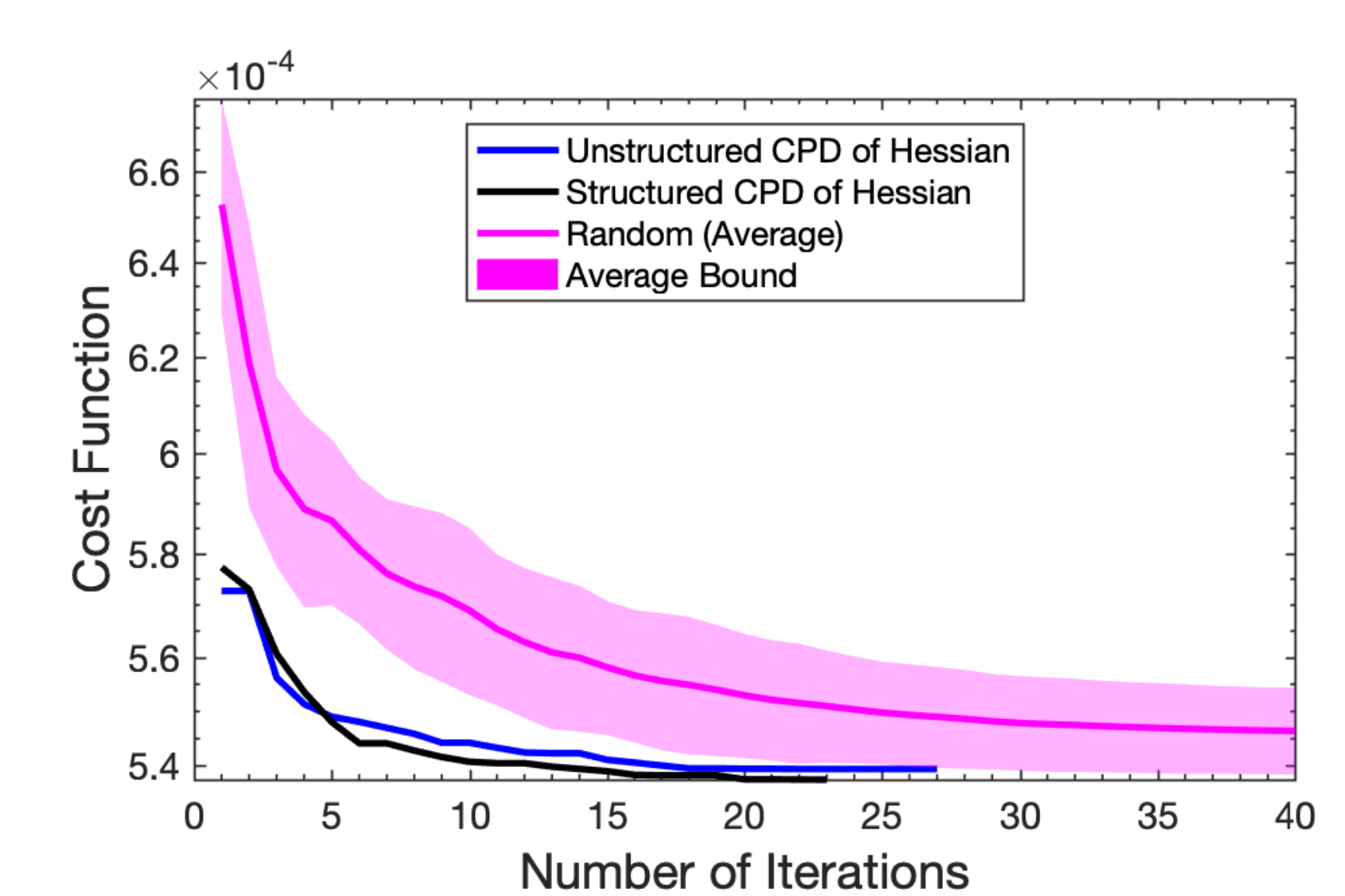}
\caption{Cost Function During the Optimization of the Bouc-Wen Model}
\label{fig:cost}
\end{figure}

Fig. \ref{fig:cost} shows the convergence of the cost function for output approximation Eq.(\ref{cost}), using different initial solutions for the optimization. It shows about a $5\% $ difference between the random and CPD initializations. This difference was also present in the validation data presented in Table \ref{tbl:BWresults}. As Fig. \ref{fig:cost} shows, it takes fewer iterations in the case of the structured CPD to reach the optimal solution in comparison to other initialization methods. However, it provides a slightly better fit than the unstructured CPD initialization. Comparing Figs. \ref{fig:Cost-SB} and \ref{fig:cost} suggests that as the system gets more complex, structural enforcement seems to play a bigger role in the optimization. In the Bouc-Wen case it took fewer steps and resulted in a lower cost function when using the structured CPD to reach the optimal value.

\begin{table}[H]
\centering
\begin{tabular}{c|c}
Initial Solution & Fit\% \\
\hline
Unstructured CPD of Hessian & 97.63\% \\
Structured CPD of Hessian & 97.63\%\\ 
Random (Average) & 97.54\% \\
\end{tabular}
\caption{Accuracy of 1-Step Ahead Prediction in the Testing Data for Various Bouc-Wen Models}
\label{tbl:BWresults}
\end{table}

\begin{figure}[h!]
\centering
\includegraphics[scale=0.5]{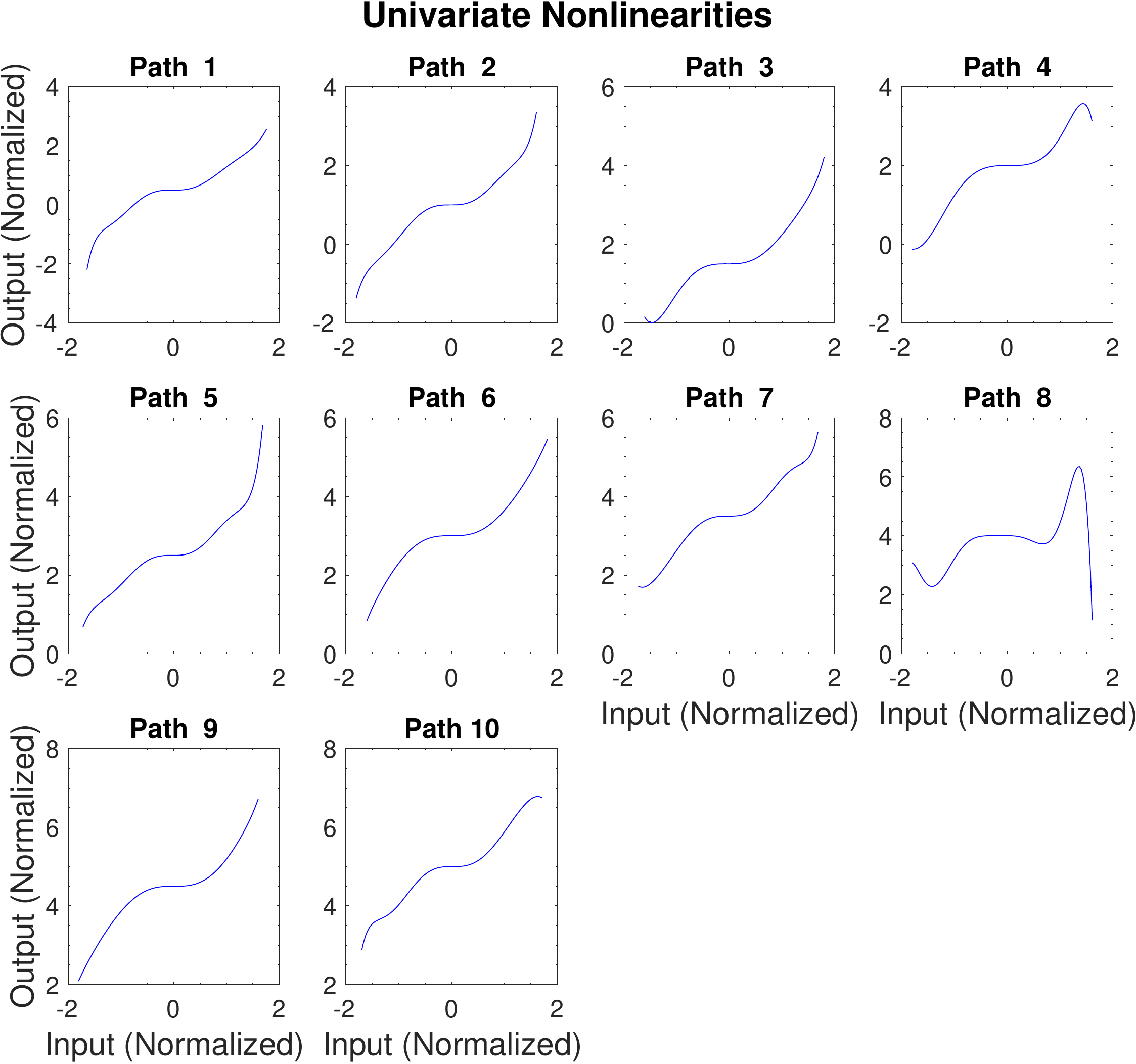}
\caption{Nonlinearities of the 10 Branch Model for the Bouc-Wen Data. The Linear Terms Have Been Removed to Emphasize the Nonlinear Aspects of the Functions}
\label{fig:UniPolyCPD2}
\end{figure}

\begin{figure}[h!]
\centering
\includegraphics[scale=0.5]{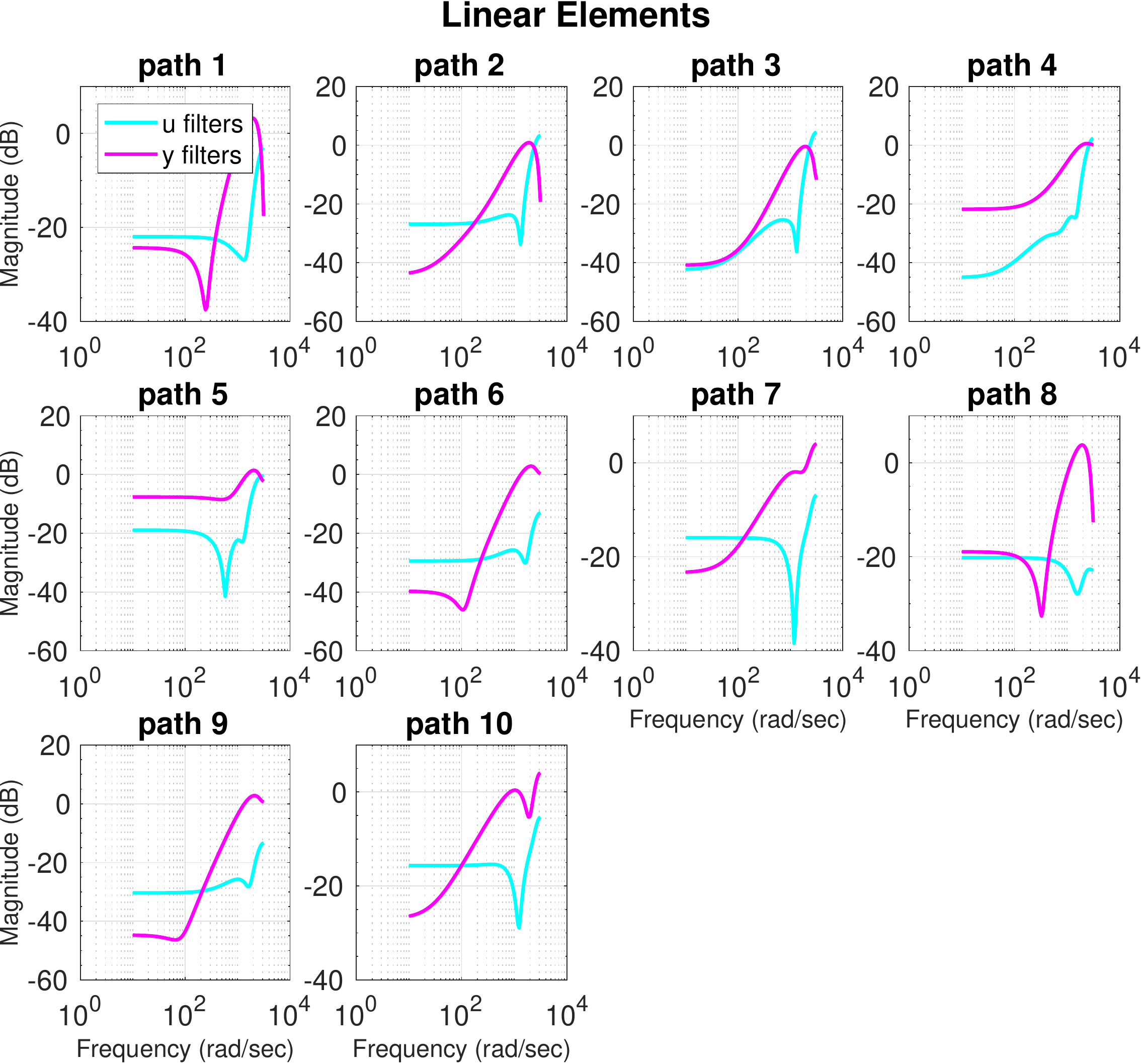}
\caption{Frequency Response Magnitudes for the Input (Cyan) and Output (Magenta) Terms of the Filters in the 10 Branches of the Bouc-Wen Models}
\label{fig:uyfilterCPD}
\end{figure}

Fig. \ref{fig:UniPolyCPD2} shows the polynomial nonlinearities in each of the $10$ branches. A dead-zone comportment is visible as all of them are flat around $0$. Figure \ref{fig:uyfilterCPD} displays the magnitude of the frequency responses of the input and output terms in the 10 filters. The input and output filters show almost the same behavior in all of the branches, the input filters have a concavity around $1500$ rad/sec and output filters show convexity around $2000$ rad/sec.

\begin{figure}[h]
\centering
\includegraphics[width = 85mm]{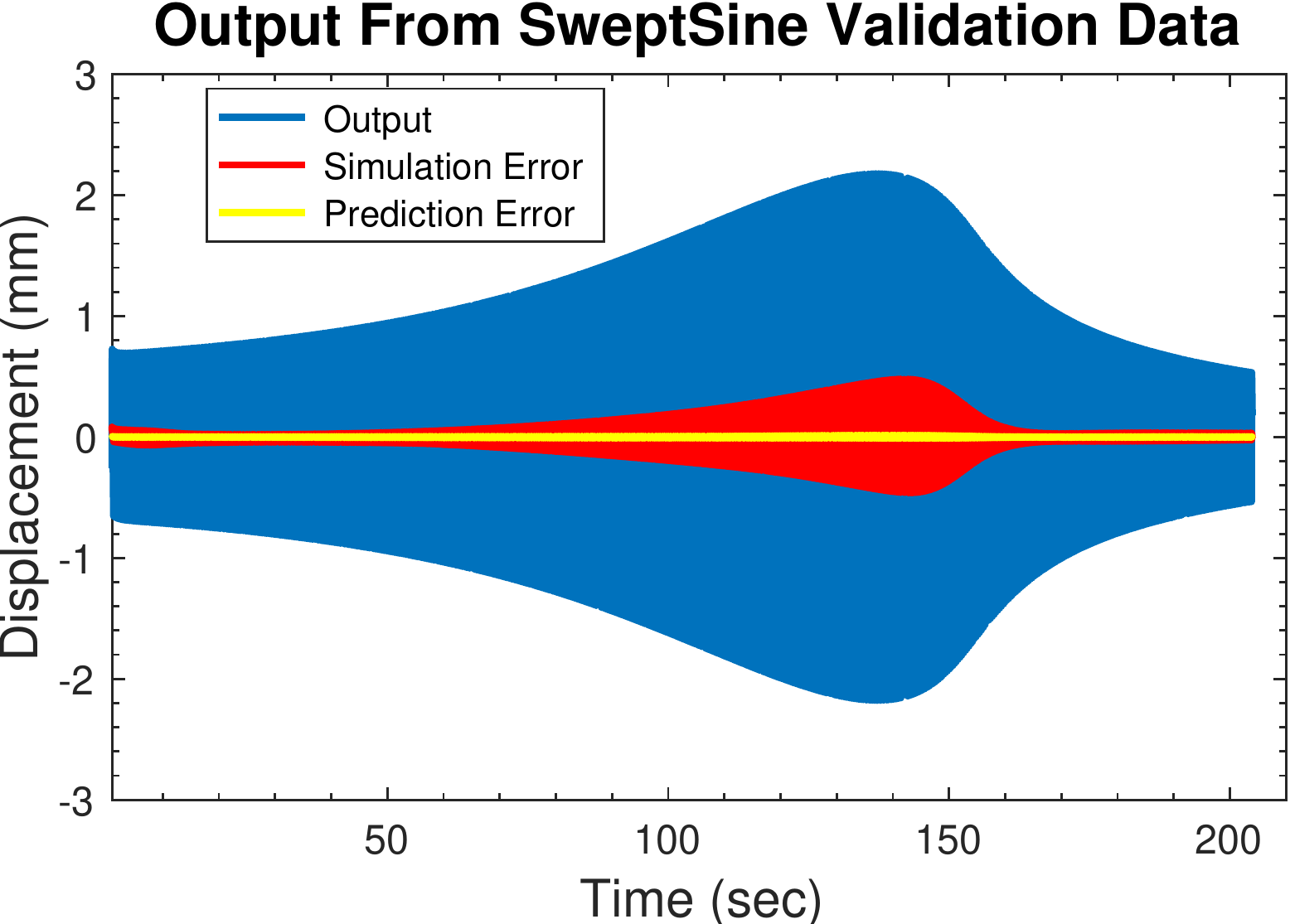}
\caption{Simulation and Prediction Errors on Swept-Sine Validation Data for Bouc-Wen}
\label{fig:BWswept}
\end{figure}

Figure \ref{fig:BWswept} shows the simulation and prediction error for swept-sine validation data set on top of the validation output data. Table \ref{tbl:BWresults5} compares the accuracy of the full P-NARX model and the decoupled P-NARX model. As illustrated in the table, the decoupled P-NARX model performs better, likely because it includes much higher degree nonlinearities while still having far fewer parameters than the full P-NARX model. Table \ref{tbl:BWresults2} includes the accuracy of the identified model using both validation data sets, the multi-sine and swept-sine. 

\begin{table}[H]
\centering
\begin{tabular}{c|c|cc}
&&\multicolumn{2}{c}{Multi-Sine }\\
 {Model}& \makecell{Number \\ of Parametersuc.}  & \makecell{Prediction\\ Accuracy}  & \makecell{Simulation\\ Accuracy} \\
\hline
Full P-NARX & 638  & 98.52\% & Unstable\\
Decouple P-NARX, 5branch & 196 & 99.92\% & 81.95\%  \\
\end{tabular}
\caption{Comparison of the Full P-NARX Model and the Decoupled P-NARX Model}
\label{tbl:BWresults5}
\end{table}

\begin{table}[H]
\centering
\begin{tabular}{c|cc|cc}
& \multicolumn{2}{c|}{Multi-Sine} & \multicolumn{2}{c}{Swept-Sine} \\ 
&  Fit\%  & $e_{RMS}$ & Fit\% & $e_{RMS}$ \\
\hline
Prediction & 99.92\%& 0.0102  & 99.60\% & 0.0037\\
Simulation & 81.95\%& 0.1705 & 85.45\% & 0.1367 \\
\end{tabular}
\caption{Accuracy of Validation Data Set for Structured CPD Initialization for Bouc-Wen Data}
\label{tbl:BWresults2}
\end{table}

\section{Conclusion} \label{conc}
The decoupled polynomial NARX structure presented in this paper, is a powerful tool to model complex nonlinear systems. This method is able to resolve major issues in the NARX model by reducing the number of parameters and providing insights of identified model for better understanding of the system. However it requires a non-convex optimization, using an appropriate initial solution will help the algorithm to achieve the optimal solution. Enforcing the polynomial constraint on one of the factors of the tensor decomposition ensures the structure of the initial solution matches that of the decoupled model. Therefore, achieving the global solution is more likely.  As the system gets more complex, the better initial solution results in getting to the optimal solution for less cost.
As mentioned in Section \ref{sec:nonpolynomial} and \ref{sec:nonPbasis}, this approach is not limited to polynomial NARX models.  Indeed, any expansion onto a smooth set of basis elements could be used for the SISO nonlinearities, provided that they retain the multiplicative structure like the polynomial basis.

\bibliography{MSEE_PhD}

\end{document}